\documentclass[aps, preprint, groupedaddress, showpacs]{revtex4}

\usepackage{amsfonts, amsbsy, amssymb, amsmath, graphicx,  float}

\usepackage{color}
\usepackage{float}
\restylefloat{figure}
\newcommand{\pd}[2]{\frac{\partial {#1}}{\partial {#2}}}

\newcommand{\calN}{\mathcal{N}}

\newcommand{\qb}{\bar{q}}
\newcommand{\pb}{\bar{p}}

\newcommand{\qdot}{\dot{q}}
\newcommand{\pdot}{\dot{p}}

\begin{document}


\title{Phase space geometry and reaction dynamics near index two saddles}



\author{Gregory S. Ezra}
\email[]{gse1@cornell.edu}
\affiliation{Department of Chemistry and Chemical Biology\\
Baker Laboratory\\
Cornell University\\
Ithaca, NY 14853\\USA}

\author{Stephen Wiggins}
\email[]{stephen.wiggins@mac.com}
\affiliation{School of Mathematics\\University of Bristol\\Bristol BS8 1TW\\United Kingdom}


\date{\today}

\begin{abstract}

We study the phase space geometry associated with index 2 saddles
of a potential energy surface and its influence on reaction dynamics for $n$ degree-of-freedom (DoF)
Hamiltonian systems. In recent years similar studies have been carried out for index 1 saddles of
potential energy surfaces, and the phase space geometry associated with classical transition state
theory has been elucidated.  In this case the existence of a normally hyperbolic invariant manifold (NHIM) of saddle stability type
has been shown, where the NHIM serves as the ``anchor''
for the construction of dividing surfaces having the no-recrossing
property and minimal flux. For the index 1 saddle case the stable and unstable manifolds of
the NHIM are co-dimension one in the energy surface
and have the structure of spherical cylinders,
and thus act as the conduits for reacting trajectories in phase space.
The situation for index 2 saddles is quite different, and their relevance for
reaction dynamics has not previously been fully recognized.
We show that NHIMs with their stable and unstable manifolds still exist, but that these manifolds
by themselves lack sufficient dimension to act as barriers in the energy surface
in order to constrain reactions. Rather, in the index 2 case there are different types of
invariant manifolds, containing the NHIM and its stable and unstable manifolds,
that act as  co-dimension one barriers in the energy surface.
These barriers divide the energy surface in the vicinity
of the index 2 saddle into
regions of qualitatively different trajectories exhibiting a wider variety of
dynamical behavior than for the case of index 1 saddles.
In particular, we can identify a class of
trajectories, which we refer to as ``roaming trajectories'', which are not associated
with reaction along the classical minimum energy path (MEP).
We illustrate the significance of our analysis of the index 2 saddle for
reaction dynamics with two examples.
The first  involves isomerization on a potential energy surface with multiple (four)
symmetry equivalent minima; the dynamics in the vicinity of the saddle enables a rigorous
distinction to be made between stepwise (sequential) and concerted (``hilltop crossing'') isomerization
pathways. The second example involves two potential minima connected by two
distinct transition states associated with conventional index one saddles, and
an index two saddle that sits between the two index one saddles.
For the case of non-equivalent index one saddles, our analysis suggests a
rigorous dynamical definition of ``non-MEP'' or ``roaming'' reactive events.

\end{abstract}


\maketitle


\section{Introduction}
\label{sec:intro}

Transition state theory has long been and continues to be a cornerstone
of the theory of chemical reaction rates \cite{Wigner38,Keck67,Pechukas81,Truhlar83,Anderson95,Truhlar96}.
In a  number of papers in recent years it has been shown that index one saddles \cite{saddle_footnote1}
of the potential energy surface
give rise to a variety of geometrical structures in {\em phase space},
enabling the realization of Wigner's vision of a transition state theory
constructed in \emph{phase space}
\cite{Wiggins90,wwju,ujpyw,WaalkensBurbanksWiggins04,WaalkensWiggins04,WaalkensBurbanksWigginsb04,
WaalkensBurbanksWiggins05,WaalkensBurbanksWiggins05c,SchubertWaalkensWiggins06,WaalkensSchubertWiggins08,
MacKay90,Komatsuzaki00,Komatsuzaki02,Wiesenfeld03,Wiesenfeld04,Wiesenfeld04a,Komatsuzaki05,Jaffe05,Wiesenfeld05,Gabern05,Gabern06,Shojiguchi08}.

Following these studies, it is natural
to investigate the phase space structure associated with saddles of index
greater than one, and to elucidate their possible dynamical significance.
In this note we describe the phase space structures and their influence on transport
in phase space  associated with  {\em index two saddles} of the potential energy surface
for $n$ degree-of-freedom (DoF) deterministic, time-independent Hamiltonian systems.

The phase space manifestation of an index one saddle of the potential energy surface
is an equilibrium point of the associated Hamilton's equations of saddle-center-$\ldots$-center stability type.
This means that the matrix associated with the linearization of Hamilton's equations about
the equilibrium point has one pair of real eigenvalues of equal magnitude, but opposite
in sign ($\pm \lambda$) and $n-1$ pairs of purely imaginary eigenvalues, $\pm i \omega_k$, $k=2, \ldots , n$.

The phase space manifestation of an index two saddle of the potential energy surface is an
equilibrium point of the associated Hamilton's equations of saddle-saddle-center-$\ldots$-center stability type.
The matrix associated with the linearization of Hamilton's equations about
the equilibrium point then has two pairs of real eigenvalues of equal magnitude, but opposite
in sign ($\pm \lambda_1, \, \pm \lambda_2$) and $n-2$ pairs of purely imaginary eigenvalues,
$\pm i \omega_k$, $k=3, \ldots , n$ \cite{saddle_footnote2}. Informally, an index two
saddle on a potential surface corresponds to a maximum or ``hilltop'' in the potential
\cite{Heidrich86,Mezey87,Wales03}.

A variety of algorithms exist for determining critical points on potential surfaces
\cite{Mezey87,Wales03,Jensen98}. The number of higher index critical points on a potential surface must be consistent
with the Morse inequalities (provided the critical points are non-degenerate)
\cite{Mezey87,Wales03}, and this constraint is often useful in ensuring that all critical points have been located.

It has however been argued that critical points of index $2$ and higher are of no
direct chemical significance \cite{Mezey87,Minyaev91}. According to the Murrell-Laidler theorem,
if two minima on a potential surface are connected by an index two saddle, then there must be a lower energy path between them
involving only ordinary transition states (index one saddles) \cite{Wales03,Murrell68}.
(There are certain well-understood limitations to the Murrell-Laidler theorem \cite{Wales92}.)
It is therefore true that, provided the reaction coordinate is defined as a minimum energy
path (MEP) \cite{Mezey87}, it cannot pass through an index two saddle on a potential surface \cite{Wales03}.
If however the reaction coordinate is instead associated with a gradient path (steepest descent/ascent)
\cite{Minyaev91,Minyaev94,Minyaev94a}, then it is
possible for an index two saddle to lie directly on the reaction path \cite{Minyaev04}.

Index two saddles (SP 2) were studied in some detail by Heidrich and Quapp \cite{Heidrich86}, who identifed two
(extreme) classes of SP 2 critical points.
The first class, ``virtual'' saddles of index two (V-SP 2), are associated
with two essentially independent index one transition states in different parts of a large molecule,
and so can be thought of simply as the direct product of two index one saddles;
an example of a V-SP 2 saddle occurs in the effectively independent internal rotation of
the methyl groups in dimethyl ether \cite{Minyaev95}.  In the second class of saddles,
``proper'' saddles of index two (P-SP 2), the motions associated with the two imaginary
vibrational frequencies at the saddle (``downhill'' motions) involve the same sets of
atoms, and so represent inherently two dimensional dynamical processes.

Examples of proper index two potential saddles abound.  Heidrich and Quapp discuss the case of
face protonated aromatic compounds, where high energy saddle points of index two prevent
proton transfer across the aromatic ring, so that proton shifts must occur at the ring periphery
\cite{Heidrich86}.  Index two saddles are found on potential surfaces
located between pairs of minima and index one saddles,
as in the case of internal rotation/inversion in the H$_2$BNH$_2$ molecule
\cite{Minyaev97}, or connected to index one saddles connecting
four symmetry related minima, as for  isomerization pathways in B$_2$CH$_4$ \cite{Fau95}.
In general, index two saddles are found separating symmetry related
transition states in so-called ``narcissistic'' transformations \cite{Salem71}.

It is quite plausible that, for low enough potential barriers \cite{Carpenter04} or at high enough energies
\cite{Meroueh02},
saddles with index $>1$ might well play a significant role in determining system properties
and dynamics.  There is for example a continuing debate concerning the role of
higher index saddles in determining the behavior of supercooled liquids and glasses
\cite{Cavagna01,Cavagna01a,Doye02,Wales03a,Angelini03,Shell04,Grigera06,Coslovich07,Angelini08}.
Computations on 256 atom LJ binary mixtures
\cite{Wales03a} show that a significant fraction of  saddles with index $\gtrsim 2$
in high-dimensional systems are however ``virtual'' saddles in the sense of Heidrich and Quapp \cite{Heidrich86}.
The general relation between configuration space topology (distribution of saddles)
and phase transitions
is also of great current interest \cite{Kastner08}.
Several examples of non-MEP or non-IRC reactions
\cite{Mann02,Sun02,Debbert02,Ammal03,Carpenter04,Lopez07,Lourderaj08} and ``roaming'' mechanisms
\cite{Townsend04,Bowman06,Shepler07,Shepler08,Suits08,Heazlewood08}
have been identified in recent years; the dynamics of these reactions is not mediated by
a single conventional transition state associated with an index one saddle.

Earnshaw's theorem \cite{Maxwell54}
suggests that higher index saddles
are likely to be ubiquitous in classical models of atoms and molecules, where
all particles interact via the Coulomb potential.
The role of index two saddles in the (classical) ionization dynamics of the Helium atom has
in fact recently been studied from a phase space perspective by Haller et al.\ \cite{haller_cite}.

Phase space structures and their influence on phase space transport
were developed in some detail for the case of an index one saddle
of the potential energy surface (henceforth, ``index one saddles'')
corresponding to an equilibrium point
of saddle-center-$\ldots$-center stability type for the corresponding Hamilton's equations \cite{ujpyw}.
In the present paper we follow a similar path for index two saddles.
(A generalization of phase space transition state theory from index one to higher index
saddles has independently been given by Haller et al.\ \cite{haller_cite}.)

The outline for this paper is as follows.
We begin in Sec.\ \ref{sec:inverted_ho}
by discussing the phase  space structure of the inverted parabolic
barrier, an index one saddle.  This elementary example provides the
foundation for our treatment of higher index saddles.
In Section \ref{sec:3dof} we consider geometry and transport associated with
a 3 DoF quadratic Hamiltonian for a saddle-saddle-center.
This simple model contains the essential features of the problem we are considering.
The quadratic Hamiltonian is separable (and integrable),
and we can therefore consider the saddle-saddle dynamics and center dynamics separately.
The saddle-saddle dynamics is considered in detail in Section \ref{sec:sad_sad}
where we show that the two integrals associated with the saddle-saddle dynamics
serve to classify the geometry of trajectories that pass through a neighborhood of the saddle-saddle.
This ``reduced dimensionality''  analysis of the saddle-saddle dynamics associated
with separability of the quadratic Hamiltonian is  central to our analysis.
We introduce a symbolic representation of the qualitatively distinct classes of
trajectory behavior in the vicinity of the saddle-saddle equilibrium, and show that
this symbolic representation can be placed into direct correspondence with the
different classes of trajectory in the vicinity of the potential hilltop.

For completeness, and in analogy with the 3 DoF case, we consider a quadratic $n$ DoF model
of a Hamiltonian having an equilibrium point
of saddle-saddle-center-$\ldots$-center equilibrium type in Appendix \ref{sec:ndof_quad},
and show that separability of the quadratic Hamiltonian implies
that the 3 DoF analysis easily goes though for the $n$ DoF case, with obvious modifications.
In Section \ref{sec:ndof} we consider the general case of a $n$ DoF fully
nonlinear Hamiltonian having an equilibrium point of saddle-saddle-center-$\ldots$-center stability type.
In a neighborhood of the equilibrium point the Poincar\'e-Birkhoff normal form theory
can be applied to construct a set of new coordinates in which the Hamiltonian assumes
a simple form (which we describe). In fact, for generic non-resonance conditions on the
eigenvalues of the matrix associated with the linearization of Hamilton's equations about
the equilibrium point, the normal form Hamiltonian is integrable. We show that integrability
provides all of the advantages that separability provided for the quadratic
Hamiltonians: the saddle-saddle dynamics can be described separately and the
two integrals associated with the saddle DoFs can be used
to characterize completely the geometry of trajectories passing through a
neighborhood of the saddle-saddle-center-$\ldots$-center equlibrium point.

As for the case of index one saddles, normally hyperbolic invariant manifolds (NHIMs)
\cite{Wiggins90,Wiggins94}
associated with index two saddles
are an important phase space structure and we describe the nature of
their existence and the role they play in phase space transport in the vicinity of
index two saddles in Sec.\ \ref{sec:compare}.

In Section \ref{sec:isomerization} we discuss two examples where phase space structure
in the vicinity of an index two saddle is potentially of importance
in a problem of chemical dynamics.
The first example involves isomerization on a potential energy surface with multiple (four)
symmetry equivalent minima; the dynamics in the vicinity of the index two saddle enables a rigorous
distinction to be made between stepwise (sequential) and concerted (``hilltop crossing'') isomerization
pathways \cite{Carpenter04}.  In the second example, two potential minima are connected by two
distinct transition states associated with conventional index one saddles, and
an index two saddle sits between the two index one saddles.
(The transition states do not have to be symmetry equivalent \cite{Salem71}.)
For high enough energies, analysis of the phase space geometry in the vicinity of
the index two saddle enables different classes of reactive trajectories to be
distinguished.  In particular, for the case of non-equivalent index one saddles, we
can rigorously define ``non-MEP'' or ``roaming'' events.
Section \ref{sec:fin_top} concludes with a discussion of some outstanding problems for future
investigation.

\newpage
\section{Index one saddle: Inverted harmonic oscillator}
\label{sec:inverted_ho}

In this section we  give a quantitative description of trajectories passing
through a neighborhood of an  equilibrium point of saddle stability type
(in brief, a ``saddle'') for 1 DoF Hamiltonian systems.
While the basic material is of course well known,
our discussion will illustrate the distinction between
physical coordinates and normal form coordinates, show how an integral
is used to distinguish qualitatively different trajectories that pass through a neighborhood of the saddle,
and develop a symbolic description that describes the qualitatively
distinct trajectories that pass through a neighborhood of the saddle.
Establishing these ideas in this simple case will allow for a straightforward
generalization to more DoF, beginning in Section \ref{sec:3dof}.

\begin{figure}[H]
\begin{center}
\includegraphics[angle=0,width=4.in]{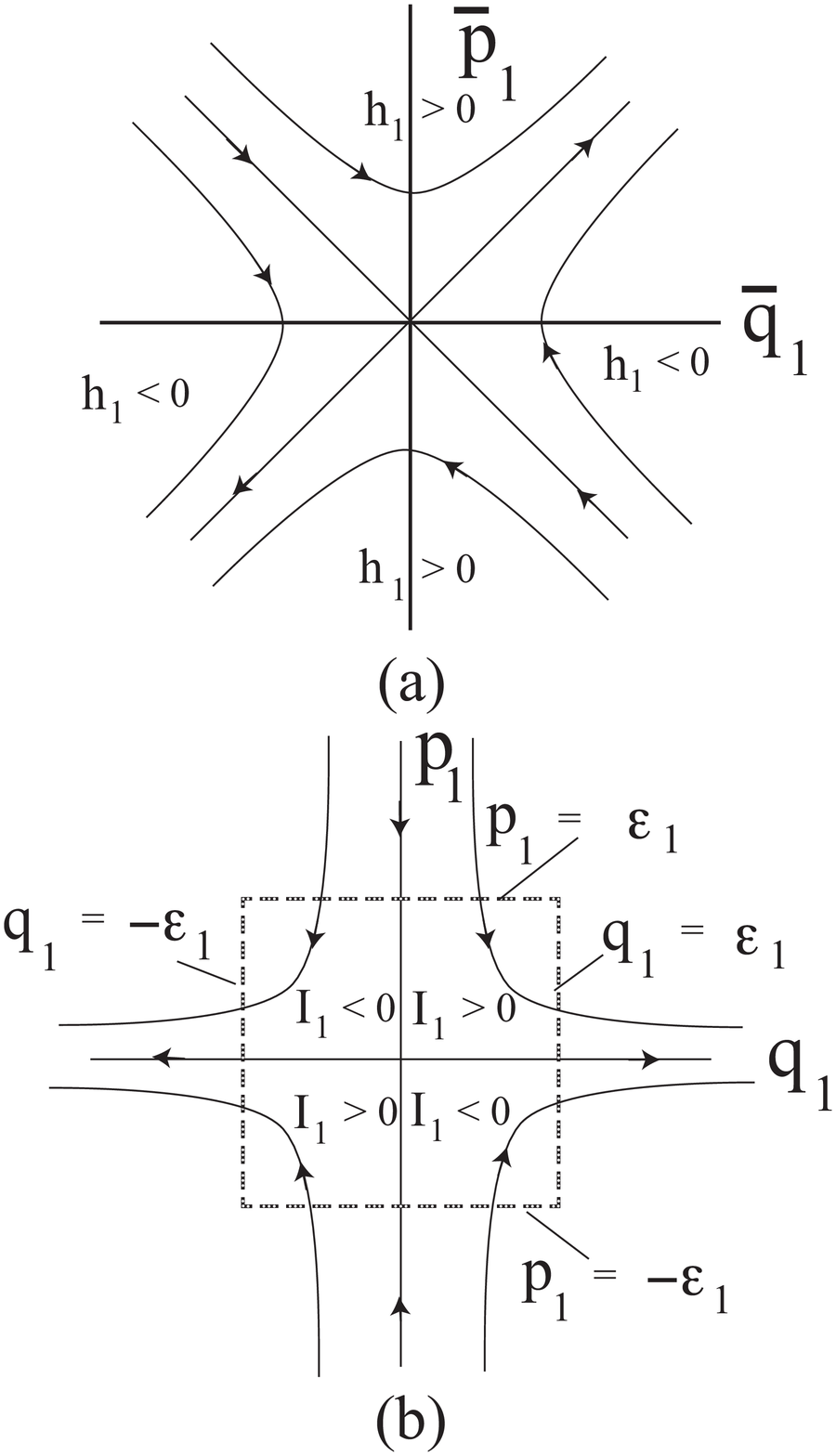}
\caption{The phase space
geometry associated with the inverted parabolic potential in (a) physical coordinates
and (b) normal form coordinates.}
\label{fig:saddle_1}
\end{center}
\end{figure}

We  consider the problem of a 1-dimensional
inverted harmonic oscillator (negative parabolic potential),
which in suitably scaled physical coordinates
$(\qb_1, \pb_1)$ corresponds to the Hamiltonian
\begin{equation}
\label{ho_1}
h_1 = \frac{\lambda_1}{2} (\pb_1^2 - \qb_1^2),
\end{equation}
with $\lambda_1 >0$.
In general, a Hamiltonian of this form provides an approximate description of
motion in the vicinity of the top of a (generic, nondegenerate)  1-dimensional barrier.
The canonical transformation $(\qb_1, \pb_1) \mapsto (q_1, p_1)$, where
\begin{subequations}
\label{ho_2}
\begin{align}
\pb_1 &= \frac{1}{\sqrt{2}} (q_1 + p_1) \\
\qb_1 &= \frac{1}{\sqrt{2}} (q_1 - p_1),
\end{align}
\end{subequations}
transforms the inverted harmonic oscillator Hamiltonian \eqref{ho_1}
into the normal form
\begin{equation}
\label{ho_3}
h_1 = \lambda_1 p_1 q_1.
\end{equation}
The associated equations of motion in normal form coordinates are
\begin{subequations}
\label{hameq_s_1}
\begin{align}
\qdot_1 & = \pd{h_1}{p_1} = \lambda_1 q_1 \\
\pdot_1 & = -\pd{h_1}{q_1} = -\lambda_1 p_1
\end{align}
\end{subequations}
with solutions
\begin{subequations}
\label{hameq_s_2}
\begin{align}
q_1 & = q_1^0 e^{+\lambda_1 t} \\
p_1 & = p_1^0 e^{-\lambda_1 t}.
\end{align}
\end{subequations}
The phase space origin $(\qb_1 =0, \pb_1 =0)$ is therefore
an equilibrium point of saddle type.
Phase space portraits for the inverted harmonic oscillator in terms of the
physical variables $(\qb_1, \pb_1)$ and the
normal form variables $(q_1, p_1)$  are shown in Figure \ref{fig:saddle_1}.
We remark that the canonical (or symplectic)  transformation to normal form coordinates,
and the Hamiltonian expressed in these coordinates (known as the ``normal form Hamiltonian'')
are trivial in this 1 DoF case.  The real power of the normal form approach becomes clear for 2 and
more DoF where it is a simple matter to describe trajectories near a higher dimensional ``saddle''
and the phase space structures near the saddle can be easily described in terms of (approximate)
integrals and the normal form coordinates (see \cite{WaalkensSchubertWiggins08} and refs therein).

The Hamiltonian \eqref{ho_3} can be written in terms of the action variable
\begin{equation}
I_1 = p_1 q_1
\end{equation}
which is obviously a conserved quantity (proportional to the Hamiltonian $h_1$).
The action $I_1$ can be positive, negative or zero.

The condition $I_1 = 0$
defines 2 co-dimension one \cite{saddle_footnote3}
invariant manifolds ($p_1 =0$ and $q_1 =0$) that intersect at the saddle equilibrium point,
which is the most basic example of a  NHIM.
More importantly, these invariant manifolds divide the phase plane into 4 regions (quadrants)
corresponding to qualitatively different types of trajectory.

In terms of the \emph{physical} variables
$(\qb_1, \pb_1)$, the two quadrants with $I_1 >0$ are associated with motions
where the particle has enough energy to surmount the top of the parabolic barrier,
so that there is no turning point in the coordinate $\qb_1$
(classical barrier transmission), whereas the
two quadrants for which the action $I_1 <0$ are associated with motions
for which the particle has insufficient energy to surmount the barrier,
so that the coordinate $\qb_1$ exhibits a turning point
(classical reflection from barrier).

To obtain a more precise classification of the possible motions in the
vicinity of the equilibrium point, we
define a phase space neighborhood $\calN$ of the equilibrium point as follows:
\begin{equation}
{\cal N} \equiv \left\{ (q_1, p_1) \, \mid \, \vert q_1 \vert \le  \varepsilon_1,
\, \vert p_1 \vert \le  \varepsilon_1 \right\},
\label{calN}
\end{equation}
\noindent
for some suitably chosen $\varepsilon_1$ \cite{saddle_footnote4}.
A trajectory enters $\calN$ if at some time
it crosses the boundary of $\calN$ with the velocity
vector pointing into $\calN$.
Similarly for trajectories that exit $\calN$. It is easy to see from \eqref{hameq_s_2} that,
with the exception of the (zero Lebesgue measure) set of trajectories that
enter $\calN$ with $\vert q_1 \vert  =0$, all trajectories that enter $\calN$,
exit $\calN$. In fact, we can characterize the condition for entry and exit of
trajectories to and from $\calN$ as follows:
\begin{description}
\item[Condition for Entry into $\cal N$:]
$\vert p_1 (0) \vert = \varepsilon_1, \, \vert I_1  \vert < \varepsilon_1^2 $.

\item[Condition for Exit from $\cal N$:] $\vert q_1 (0) \vert = \varepsilon_1,
\, \vert I_1  \vert < \varepsilon_1^2 $.

\end{description}

Trajectories that exit and enter $\calN$ remain in the same quadrant in the $(q_1, p_1)$
plane, and either pass over or are reflected by the potential barrier
(this latter interpretation is based on the Hamiltonian expressed in terms of the physical coordinates).
Corresponding to the four distinct quadrants, we define four classes of trajectory:
\begin{itemize}

\item $(++)$ ($q_1>0$, $p_1<0$)

\item $(-+)$ ($q_1<0$, $p_1<0$)

\item $(+-)$ ($q_1>0$, $p_1>0$)

\item $(--)$ ($q_1<0$, $p_1>0$)

\end{itemize}

This description uses both the physical and normal form coordinates,
and the notation indicates the net result of passage through region $\calN$
in terms of the action of an associated ``scattering map''.
Considering the first item, reading from right to left \cite{qm_footnote}, $(+,+)$ refers to a
trajectory in the physical coordinates that enters the neighborhood of the saddle with $\qb_1 >0$
and exits the neighborhood of the saddle  with $\qb_1 >0$, i.e. it is reflected
from the barrier on the $\qb_1 >0$ side. In the normal form coordinates this trajectory corresponds
to a trajectory in the fourth quadrant, ($q_1>0$, $p_1<0$). Similarly,
trajectories of type $(-+)$ enter a neighborhood of the saddle with $\qb_1> 0$,
pass over the saddle, and exit the neighborhood of the saddle with  $\qb_1<0$.
In normal form coordinates these correspond to trajectories  in the third quadrant,
($q_1<0$, $p_1<0$). The remaining  two cases are understood similarly.

\newpage
\section{Phase Space Geometry and Transport Associated with a 3 DoF Quadratic Hamiltonian of
a Saddle-Saddle-Center Equilibrium}
\label{sec:3dof}

In this section we describe the nature of trajectories as they pass through a
neighborhood of an index 2 saddle for a 3 DoF  quadratic  Hamiltonian.
We will express the Hamiltonian in normal form coordinates.  However, the
symbolic representation of the trajectories will be based on dynamics described
in terms of the
physical coordinates, as in the previous section.

\subsection{Quadratic saddle-saddle-center system}
\label{sec:3dof_quad}

We now consider a  3 DoF quadratic Hamiltonian that describes the dynamics near a
saddle-saddle-center  stability type equilibrium point located at the origin:
\begin{equation}
H_{ssc} = \lambda_1 q_1 p_1 +  \lambda_2 q_2 p_2 +
\frac{\omega}{2} \left( q_3^2 + p_3^2 \right), \quad \lambda_1, \, \lambda_2, \, \omega >0.
\label{model_lin_ham_1}
\end{equation}
Note that the variables $(q_1, p_1, q_2, p_2)$ in Hamiltonian \eqref{model_lin_ham_1}
are not physical coordinates, but rather normal form coordinates
as defined in eq.\ \eqref{ho_2}.  The Hamiltonian \eqref{model_lin_ham_1}
can be used to describe dynamics in the vicinity of a hill-top (nondegenerate index two saddle) in a
3 dimensional potential energy surface.

Hamiltonian \eqref{model_lin_ham_1} is completely integrable, with (independent) integrals:
\begin{equation}
I_1 = q_1 p_1, \, I_2 = q_2 p_2, \, I_3 = \frac{1}{2} \left( q_3^2 + p_3^2 \right),
\label{int_3dof}
\end{equation}
\noindent
and can be expressed as a function of the integrals:
\begin{equation}
H_{ssc} = \lambda_1 I_1 +  \lambda_2 I_2 + \omega I_3.
\label{ham_int_3dof}
\end{equation}
\noindent
These integrals play a crucial role in our understanding of the underlying phase space geometry and transport.

\subsection{Phase Space Geometry and Transport Associated with the 2 DoF Saddle-Saddle Subsystem}
\label{sec:sad_sad}

An important consequence of the separability of the 3 DoF quadratic Hamiltonian
\eqref{model_lin_ham_1} is
that an understanding of the nature of the trajectories passing through a neighborhood
of the saddle-saddle-center equilibrium point can be obtained through a study of the 2 DoF
subsystem corresponding to the saddle-saddle, since motion in the  center DoF is bounded,
and the action $I_3$ is conserved.
The Hamiltonian for this 2 DoF subsystem is:
\begin{equation}
H_{ss} = \lambda_1 q_1 p_1 +  \lambda_2 q_2 p_2,
\label{model_lin_ham_saddle}
\end{equation}
\noindent
with integrals $I_1$ and $I_2$ as given in \eqref{int_3dof}.

\begin{figure}[H]
\begin{center}
\includegraphics[angle=270,width=5.5in]{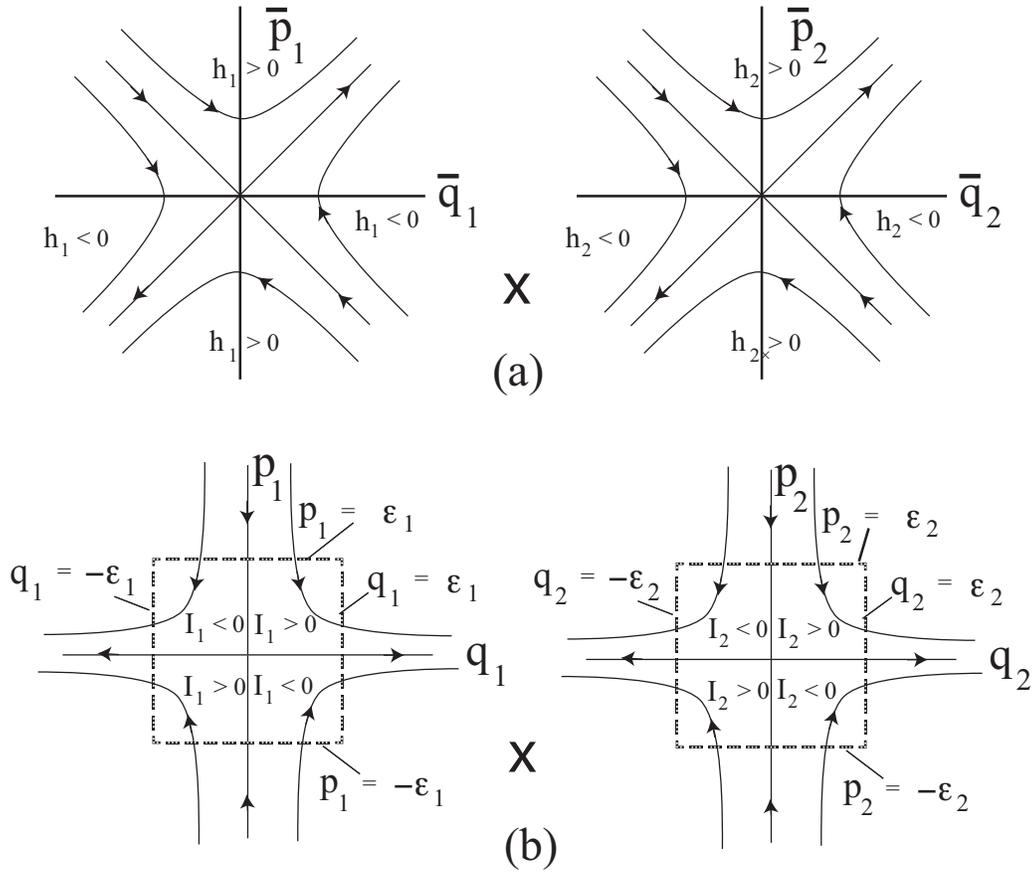}
\caption{The geometry associated with the 2 DoF saddle-saddle subsystem.}
\label{fig:saddle_saddle}
\end{center}
\end{figure}

Hamilton's equations for the 2 DoF saddle-saddle subsystem are given by:
\begin{subequations}
\label{hameq_ss_1}
\begin{align}
\dot{q}_1 & =  \frac{\partial H_{ss}}{\partial p_1}= \lambda_1 q_1,  \\
\dot{p}_1 & =   -\frac{\partial H_{ss}}{\partial q_1}=  -\lambda_1 p_1, \\
\dot{q}_2 & =   \frac{\partial H_{ss}}{\partial p_2}=  \lambda_2 q_2,  \\
\dot{p}_2 & =  -\frac{\partial H_{ss}}{\partial q_2}=  -\lambda_2 p_2
\end{align}
\end{subequations}
The phase flow in the vicinity of the saddle-saddle equilibrium is
shown in Fig.\ \ref{fig:saddle_saddle}.  In this figure we show the
phase flow in both the physical coordinates defined in \eqref{ho_1}
and the normal form coordinates. The  transformation between these two sets of
coordinates is given in \eqref{ho_2}. Since our Hamiltonian is separable, it is a
trivial matter to use this coordinate transformation on both saddles to express the Hamiltonian
in physical coordinates. However, the separability also means that we can immediately
adapt the discussion of the previous section for physical coordinates to the saddle-saddle case,
and it is this approach that we will follow.

As before, we define a neighborhood $\calN$ of the
equilibrium point in phase space as follows:
\begin{equation}
{\cal N} \equiv \left\{ (q_1, p_1, q_2, p_2) \, \mid \, \vert q_1 \vert \le  \varepsilon_1,
\, \vert p_1 \vert \le  \varepsilon_1, \,
\vert q_2 \vert \le  \varepsilon_2, \, \vert p_2 \vert \le  \varepsilon_2
\right\},
\label{calB}
\end{equation}
\noindent
for some suitably chosen $\varepsilon_1$ and $\varepsilon_2$ \cite{saddle_footnote4}.
As for the 1 DoF case,
with the exception of the (zero Lebesgue measure) set of trajectories that
enter $\cal N$ with $\vert q_1 \vert = \vert q_2 \vert =0$, all trajectories that enter $\cal N$,
exit $\cal N$.   We characterize the condition for entry and exit of
trajectories to and from $\cal N$ as follows:
\begin{description}

\item[Condition for Entry into $\cal N$:]
$\vert p_1 (0) \vert = \varepsilon_1, \, \vert p_2 (0) \vert = \varepsilon_2, \,
\vert I_1 I_2 \vert < \varepsilon_1^2 \varepsilon_2^2$.

\item[Condition for Exit from $\cal N$:] $\vert q_1 (0) \vert = \varepsilon_1,
\, \vert q_2 (0) \vert = \varepsilon_2,
\, \vert I_1 I_2 \vert < \varepsilon_1^2 \varepsilon_2^2$.

\end{description}

As for the 1 DoF unstable equilibrium, we can define
different classes of trajectory characterized by their behavior under
the scattering map corresponding to passage through
the region $\calN$.  For the 2 DoF case, there are $4 \times 4 = 16$ combinations
of quadrants, and $16$ types of trajectory. The symbolic description of the behavior
of a trajectory as it passes through a neighborhood of the saddle-saddle with respect
to the physical coordinates $\qb_k, \, k=1, 2$, is expressed by the following four symbols,
$(f_1 f_2; i_1 i_2)$,
where $i_1 = \pm$, $i_2 = \pm$, $f_1 = \pm$, $f_2 = \pm$.
Here $i_k, \, k=1,2$ refer to the ``initial'' sign of $\qb_k, \, k=1,2$
as it enters the neighborhood of the saddle-saddle and
$f_k, \, k=1,2$ refer to the ``final'' sign of $\qb_k, \, k=1,2$,
as it leaves the neighborhood of the saddle.
For example, trajectories of class $(--;+-)$ pass over the
barrier from $\qb_1 >0$ to $\qb_1 <0$, but remain on the side
of the barrier with $\qb_2 <0$.

Of the 16 qualitatively distinct classes of trajectory, the 4 types
$(++;--)$, $(-+;+-)$, $(+-;-+)$ and $(--;++)$ are the
only trajectories that undergo a change of
sign of both coordinates $\qb_1$ and $\qb_2$,
and so they are the trajectories that  pass ``over the hilltop''
in the vicinity of the saddle-saddle equilibrium.


As in the 1 DoF case, co-dimension one surfaces separate the different types
of trajectory.  These are the four co-dimension one invariant manifolds
given by $q_1=0, \, p_1 =0, \, q_2 =0, \, p_2 =0$ (i.e. $I_1 =0, \, I_2 =0$).
For example, the co-dimension one surface $q_1 = 0$
forms the boundary between trajectories of type $(++; +-)$ and
$(-+; +-)$, and so on.

The dynamical significance of this symbolic classification is disussed in
Section \ref{sec:isomerization} in the context of isomerization reactions.

\subsection{Including the Additional Center DoF}
\label{sec:sad_sad_cent}

Because the system is separable, including the additional center DoF has no effect on our discussion above
on the nature of trajectories in the 2 DoF saddle-saddle subsystem as they pass through
a neighborhood of the saddle.

To see this, note that equations of motion for Hamiltonian \eqref{model_lin_ham_saddle} are given by:
\begin{subequations}
\label{hameq_ss}
\begin{align}
\dot{q}_1 & =  \frac{\partial H_{ssc}}{\partial p_1}   = \lambda_1 q_1,  \\
\dot{p}_1 & =   -\frac{\partial H_{ssc}}{\partial q_1}  =  -\lambda_1 p_1,  \\
\dot{q}_2 & =  \frac{\partial H_{ssc}}{\partial p_2}   =  \lambda_2 q_2,  \\
\dot{p}_2 & =   -\frac{\partial H_{ssc}}{\partial q_2}  =  -\lambda_2 p_2, \\
\dot{q}_3 & =   \frac{\partial H_{ssc}}{\partial p_3}   =  \omega p_3,  \\
\dot{p}_3 & =  -\frac{\partial H_{ssc}}{\partial q_3}   =  -\omega q_3,
\end{align}
\end{subequations}
\noindent
As previously, we define a neighborhood of the saddle-saddle-center
equilibrium point in phase space as follows:
\begin{equation}
{\cal N} \equiv \left\{ (q_1, p_1, q_2, p_2, q_3, p_3) \, \mid \, \vert q_1 \vert \le  \varepsilon_1, \, \vert p_1 \vert \le  \varepsilon_1, \, \vert q_2 \vert \le  \varepsilon_2, \, \vert
p_2 \vert \le  \varepsilon_2 \,
\vert q_3 \vert \le  \varepsilon_3, \, \vert p_3 \vert \le  \varepsilon_3
\right\},
\label{calN_2}
\end{equation}
\noindent
for suitably chosen $\varepsilon_1$, $\varepsilon_2$ and $\varepsilon_3$.
We want to describe the geometry associated with trajectories that enter and leave $\cal N$.
First, note again the fact that Hamilton's equations are separable, and that the
(simple harmonic) motion of $q_3$ and $p_3$ is bounded.
It then follows that if $q_3$ and $p_3$ are initially chosen to satisfy the condition
for being in $\cal N$,
then they satisfy this condition for all time. Hence, the issue of trajectories
entering $\cal N$ and exiting $\cal N$ depends entirely on the behavior of the $(q_1, p_1, q_2, p_2)$
components of a trajectory and, as a consequence of separability,
this behavior is exactly as described in Section \ref{sec:sad_sad}.

We now examine some further aspects of the geometry for the 3 DoF quadratic Hamiltonian
model of a saddle-saddle-center. The equilibrium point is located at the origin and has zero energy.
We consider geometric structures in the energy surface for positive energies:
\begin{equation}
H_{ssc} = \lambda_1 q_1 p_1 +  \lambda_2 q_2 p_2 + \frac{\omega}{2} \left( q_3^2 + p_3^2 \right)=E>0.
\label{E_surf_3dof}
\end{equation}
\noindent
It is clear from \eqref{hameq_ss} that $q_1=p_1=q_2=p_2=0$ is a two dimensional
invariant manifold in the 6 dimensional phase space. Using \eqref{E_surf_3dof},
its intersection with the 5 dimensional energy surface is given by:
\begin{equation}
\frac{\omega}{2} \left( q_3^2 + p_3^2 \right)=E>0.
\label{nhim_3dof}
\end{equation}
\noindent
This is a periodic orbit in the energy surface but, more generally, it is an example of a NHIM.
The coordinates $q_1, p_1, q_2, p_2$ can be viewed as coordinates for the normal directions of the NHIM,
and it follows from the nature of the asymptotic (in time) behavior of these coordinates
(see \eqref{hameq_ss}) that this NHIM has a 3 dimensional stable manifold and a 3 dimensional
unstable manifold in the 5 dimensional energy surface.
Further discussion of the role of NHIMs is given in Section \ref{sec:nhim}.

\newpage

\section{$n$ DoF, Higher Order Terms in the Hamiltonian, and the Poincar\'e-Birkhoff Normal Form}
\label{sec:ndof}

The examples we have considered so far have been exceptional -- quadratic,
separable -- Hamiltonians (which implies that they are completely integrable).
Now we will show that  for  a general time-independent Hamiltonian, in a neighborhood
of an equilibrium point of  saddle-saddle-center-$\ldots$-center stability type,
a ``good set of coordinates'' can be found
in terms of which the Hamiltonian assumes a simple form for which the results described above hold.
Moreover, these coordinates are obtained via a  constructive algorithm.

For completeness, in Appendix \ref{sec:ndof_quad} we consider in detail
the nature of trajectories associated with a quadratic $n$ DoF Hamiltonian model that describes
the dynamics near a saddle-saddle-center-$\ldots$-center stability type equilibrium point,
and then we consider the case where higher order terms are included in this model.

\subsection{Normal form for $n$ DoF}

Specifically, in a neighborhood of  an equilibrium point of
saddle-saddle-center-$\ldots$-center stability type, Poincar\'e-Birkhoff normal
form theory can be used to (formally) construct a symplectic change of coordinates
so that in the new coordinates the Hamiltonian has the form:
\begin{equation}
H = \lambda_1 q_1 p_1 +  \lambda_2 q_2 p_2 +
\sum_{i=3}^{n} \frac{\omega_i}{2} \left( q_i^2 + p_i^2 \right) + f(q_1, \ldots, q_n, p_1, \ldots , p_n),
\label{model_nonlin_ham}
\end{equation}
\noindent
where $f(q_1, \ldots, q_n, p_1, \ldots , p_n)$ assumes
a particularly simple form that is amenable to analysis. This is discussed in some detail,
with particular relevance to reaction dynamics, in ref.\ \cite{WaalkensSchubertWiggins08}.
For our purposes, if we assume that the purely imaginary eigenvalues satisfy
the non-resonance condition $k_3 \omega_3 + \ldots + k_n \omega_n \ne 0$
for any $n-2$ vector of integers $(k_3, \ldots, k_n)$ with {\em not all}
the $k_i=0$ (that is, $(k_3, \ldots , k_n) \in {\mathbb Z}^{n-2}-\{0\}$)
and the real
eigenvalues satisfy the (independent) non-resonance condition $k_1 \lambda_1 +  k_2 \lambda_2 \ne 0$
for any $2$ vector of integers $(k_1,  k_2)$ with {\em not all}
the $k_i=0$, $i=1,2$, then $f(q_1, \ldots, q_n, p_1, \ldots , p_n)$ can be represented as an even order polynomial in the
variables
\begin{equation}
I_1 = q_1 p_1, \, I_2 = q_2 p_2, \, I_k = \frac{\omega_k}{2} \left( q_k^2 + p_k^2 \right), \, k=3, \ldots, n.
\label{ndof_ints}
\end{equation}

In other words, we can express the normal form Hamiltonian in this situation as \cite{normal_form_footnote1}:
\begin{equation}
H(I_1, I_2, I_3, \ldots, I_n),
\label{nf_ham}
\end{equation}
\noindent
with the associated Hamilton's equations given by:
\begin{subequations}
\label{nf_hameq}
\begin{align}
\dot{q}_1 & =  \frac{\partial H}{\partial p_1}= \frac{\partial H}{\partial I_1}\frac{\partial I_1}{\partial p_1} =  \frac{\partial H}{\partial I_1} q_1,  \\
\dot{p}_1 & =  -\frac{\partial H}{\partial q_1}= -\frac{\partial H}{\partial I_1}\frac{\partial I_1}{\partial q_1}=  -\frac{\partial H}{\partial I_1} p_1, \\
\dot{q}_2 & =  \frac{\partial H}{\partial p_2}= \frac{\partial H}{\partial I_2}\frac{\partial I_2}{\partial p_2} =  \frac{\partial H}{\partial I_2} q_2,  \\
\dot{p}_2 & =  -\frac{\partial H}{\partial q_2}= -\frac{\partial H}{\partial I_2}\frac{\partial I_2}{\partial q_2}=  -\frac{\partial H}{\partial I_2} p_2,  \\
\dot{q}_k & =  \frac{\partial H}{\partial p_k}= \frac{\partial H}{\partial I_k}\frac{\partial I_k}{\partial p_k} =  \frac{\partial H}{\partial I_k} p_k,  \\
\dot{p}_k & =  -\frac{\partial H}{\partial q_k}= -\frac{\partial H}{\partial I_k}\frac{\partial I_k}{\partial q_k}=  -\frac{\partial H}{\partial I_k} q_k,
\qquad k=3, \ldots, n,
\end{align}
\end{subequations}
\noindent
and it can be verified by a direct calculation that the $I_k$, $k=1, \ldots, n$ are
integrals of the motion for \eqref{nf_hameq} \cite{saddle_footnote5}.

Since the $I_k$, $k=1, \ldots, n$ are constant on trajectories,
it follows that the partial derivatives $\frac{\partial H}{\partial I_k}, \, k=1, \ldots, n,$ are
also constant on trajectories. Hence, it follows that the structure of eqs \eqref{nf_hameq}
is extremely simple, and this is a consequence of the integrability of the normal form.

We can define a neighborhood of the equilibrium point exactly as in eq.\ \eqref{calL}
in Appendix \ref{sec:ndof_quad} and consider
the geometry associated with trajectories that enter and leave $\cal L$.
Even though the nonlinear Hamilton's equations \eqref{nf_hameq} are {\em not} separable,
it follows from integrability and the structure of \eqref{nf_hameq}
that the motion of $q_k$ and $p_k$ is bounded, $k=3, \ldots, n$  ( $q_k$ and $p_k$
just undergo periodic motion, individually for each $k$, although the
frequency of motion can vary from trajectory to trajectory).
It then follows that if $q_k$ and $p_k$, $k=3. \ldots, n$, are initially chosen in $\cal L$,
then they remain in $\cal L$ for all time. Hence, exactly as in the quadratic case,
the issue of trajectories entering $\cal L$ and exiting $\cal L$ depends
entirely on the behavior of the $(q_1, p_1, q_2, p_2)$ components of a trajectory and,
as a consequence of integrability,  this behavior is exactly as described in Section \ref{sec:sad_sad}
and we have the integrals $I_1$ and $I_2$ as parameters for describing the geometry of the
trajectories during their passage though $\cal L$.

Exactly as for the  model quadratic Hamiltonians, it is simple to verify that $q_1=p_1=q_2=p_2=0$
is a $2n-4$ dimensional invariant manifold in the $2n$ dimensional phase space.
For energies greater than zero (the energy of the saddle-saddle-center-$\ldots$-center),
in the $2n-1$ dimensional energy surface it is a $2n-5$ dimensional NHIM with $2n-3$
dimensional stable and unstable manifolds
(they are co-dimension two \cite{saddle_footnote3}
 in the energy surface).
 An important property of NHIMs is that they, as well as their stable and unstable manifolds,
 persist under perturbation. If we consider a small $\cal L$,
 then the terms $f(q_1, p_1, \ldots, q_n, p_n)$ can be considered as
 a perturbation of the quadratic Hamiltonian, for
 which the NHIM had the geometrical structure of $S^{2n-5}$.
 If the perturbation is ``sufficiently small'' then this spherical structure is preserved.

 \subsection{Accuracy of the Normal Form}
 \label{sec:accuracy}

The normalization proceeds via
formal power series manipulations whose input is a Taylor
expansion of the original Hamiltonian, $H$, necessarily up to some
finite order, $M$, in homogeneous polynomials.  For a particular
application, this procedure naturally necessitates a suitable
choice of the order, $M$, for the normalization, after which one
must make a restriction to some local region, $\cal L$, about
the equilibrium point in which the resulting computations achieve
some desired accuracy. Hence, the accuracy of the normal form as a power series expansion truncated
at order $M$ in a neighborhood $\cal L$ is determined by comparing the dynamics associated with
the normal form with the dynamics of the original system. There are several independent tests
that can be carried out to verify accuracy of the normal form. Straightforward tests
that we have used are the following
\cite{WaalkensBurbanksWiggins04,WaalkensBurbanksWigginsb04,WaalkensBurbanksWiggins05,WaalkensBurbanksWiggins05b,
WaalkensBurbanksWiggins05c,WaalkensSchubertWiggins08}:
\begin{itemize}

\item Examine how the integrals associated with the normal form change on
trajectories of the full Hamiltonian (the integrals will be constant on trajectories of the normal form).

\item Check invariance of the different invariant manifolds
(i.e. the NHIM and its stable and unstable manifolds) with respect to trajectories of the full Hamiltonian.

\end{itemize}

Both of these tests require us to use the transformations between the original coordinates and the normal form coordinates.
Specific examples where $M$, $\cal L$ and accuracy of the
normal forms and the constancy of integrals of the truncated normal form
are considered can be found in
\cite{WaalkensBurbanksWiggins04,WaalkensBurbanksWigginsb04,WaalkensBurbanksWiggins05,WaalkensBurbanksWiggins05b,WaalkensBurbanksWiggins05c}.
A general discussion of accuracy of the normal form can be found in \cite{WaalkensSchubertWiggins08}.

\newpage

\section{Comparison of Phase Space Geometry and Trajectories for Index One and Index Two Saddles}
\label{sec:compare}

In this section we discuss a number of issues that arise from the above results and discussions.
For comparative purposes it is helpful first to summarize the nature of trajectories that pass near
an index one saddle and trajectories that pass near an index two saddle.

\subsection{Index one saddles}

For index one saddles it was shown in previous work that for energies
above that of the saddle-center-$\ldots$-center there exists a
$2n-3$ dimensional normally hyperbolic invariant manifold (NHIM) in the
$2n-1$ dimensional energy surface \cite{Wiggins90,Wiggins94,ujpyw}. The NHIM is of saddle stability type and has
$2n-2$ dimensional stable and unstable manifolds.
Since these are one less dimension than the energy surface,
they act as higher dimensional separatrices and serve to
partition phase space into regions corresponding to qualitatively different types of trajectories:
\begin{itemize}

\item Forward reactive trajectories $\equiv (+-)$

\item Forward non-reactive trajectories $\equiv (--)$

\item Backward reactive trajectories $\equiv (-+)$

\item Backward non-reactive trajectories $\equiv (++)$

\end{itemize}
where we have indicated the correspondence with the trajectory classification introduced in Sec.\
\ref{sec:inverted_ho}.

The NHIM is the ``anchor'' for the construction of a phase space
dividing surface having the geometrical structure of a $2n-2$ dimensional sphere.
The NHIM is the equator of this $2n-2$ dimensional sphere, hence dividing
it into two hemispheres. Forward reacting trajectories pass through one
hemisphere and backward reacting trajectories pass through the other hemisphere.
These ``dividing hemispheres'' have the \emph{no-recrossing property} in the sense that
the vector field defined by Hamilton's equations is strictly transverse to each hemisphere.
Moreover, the magnitude of the flux is minimal (and equal) for each hemisphere.

The language used for the description of trajectories for the case of an index
one saddle is that of reaction. Trajectories evolve from reactants to products
by passing through a dividing surface that locally ``divides'' the energy surface into
two disjoint pieces--reactants and products.

Here, we have classified trajectories in the vicinity of an index one saddle
in a slightly less general but more explicit fashion
in terms of their behavior with respect to crossing a potential barrier; this classification
is useful in the treatment of reactive dynamics in the vicinity of
higher index saddles on potential energy surfaces, as discussed below.

\subsection{Index two Saddles}

The geometrical structures associated with index one saddles provide a rigorous way of
partitioning the phase space into regions
that describe the evolution of a system from reactants to products by passing through
a dividing surface having the no-recrossing and minimal flux properties.
This partitioning is made possible by the existence of invariant manifolds
that have the proper dimension and stability type. Moreover, their geometric character is such that the partitioning that they provide
has the natural interpretation  of regions of reactants and products of the energy surface, and the stable and unstable manifolds of the NHIM provide natural boundaries between these
regions.

For index two saddles
the same types of invariant manifolds still exist -- a NHIM and co-dimension one
(in the energy surface) invariant manifolds. These co-dimension
one invariant manifolds ($q_i = 0$, $p_i=0$, $i=1,2$) were
introduced above and are discussed further below.  It was shown that the passage of trajectories
through the neighborhood of the equilibrium could be understood in terms of
a 2 DoF saddle-saddle subsystem and values of the two associated action  integrals.

For index one saddles not all trajectories could pass through the dividing surface
(by definition, the forward and backward non-reactive trajectories do not do so).
However, for index two saddles all
trajectories (except for the set of Lebesgue measure zero discussed above) that enter
the neighborhood of the equilibrium point exit the neighborhood of the equilibrium point,
and there are sixteen qualitatively different types of trajectory
that enter and exit this neighborhood.
These sixteen classes are characterized by their crossing behavior with respect to
the ``hilltop'' in the potential surface associated with the index two saddle.

Two examples involving isomerization reactions
where this classification of trajectories is of chemical relevance are
discussed in  Section \ref{sec:isomerization}.

\subsection{The Role of NHIMs and the existence of co-dimension one invariant manifolds in the energy surface}
\label{sec:nhim}

It has been previously been argued \cite{ujpyw} that the invariant manifold structure associated with
index one saddles is a natural generalization to 3 or more DoF of the PODS,
and their associated invariant manifolds, where the latter have provided
a fundamental understanding of transition state theory for systems with
2 DoF  \cite{Pechukas81}.
The necessary ingredient to make the leap from 2 DoF to higher DoF is the
realization that a NHIM of saddle stability type is the natural  generalization
to higher DoF of the periodic orbit of saddle stability type familiar from 2 DoF \cite{Pollak78,ujpyw}.
The resulting geometric picture of reaction for $N > 2$ DoF is essentially
that of the 2 DoF case, the key feature of the generalization being the
availablity of the necessary invariant manifolds to partition the energy surface in the multidimensional case in a manner that is meaningful for reaction.

The situation with index two saddles is fundamentally different.
On energy surfaces above that of the saddle,  a NHIM  still exists,
but it is co-dimension 4 in the energy surface with co-dimension 2 stable and
unstable manifolds. Nevertheless, as shown above,
it is still possible to  construct co-dimension one
invariant manifolds (in the energy surface) that locally
partition the energy surface into
regions associated with qualitatively different types of trajectory
passing through a neighborhood of the equilibrium point:
the surfaces defined by the conditions
$q_1=0$, $p_1=0$ $q_2=0$, and $p_2=0$ are each
co-dimension one invariant manifolds in the $2n$ dimensional phase space,
and they are also co-dimension one invariant manifolds when intersected with the energy surface.
They do not have the property of being compact and boundaryless like the NHIMs we have
discussed above. We would not expect this since (almost) all trajectories that enter $\cal L$ leave $\cal L$.
These co-dimension one invariant manifolds intersect the boundary of $\cal L$,
and their projection into the $q_1-p_1-q_2-p_2$ space gives the
coordinate axes shown in Fig \ref{fig:saddle_saddle}.
This is consistent with our description of trajectories that enter
and exit $\cal L$ through the use of the integrals $I_1$ and $I_2$.

While the existence of  co-dimension one invariant manifolds in phase space associated
with index one or index two saddles is conceptually appealing,  practically, their main use
derives from the integrable structure of the normal form and the expression of Hamilton's
equation using this integrability (e.g. \eqref{nf_hameq}). This expression allows for
a decoupling of the dynamics into unbounded motion (the ``reactive modes'' described by $I_1$ and $I_2$)
and bounded motion (the ``bath modes'' described by $I_k, \, k=3, \ldots, n$).
The ``reactive dynamics'' is described by a reduced dimensional system and the
dynamics in this system is integrable and characterized by the level sets of these integrals.
The relevant co-dimension one invariant manifolds drop out of this description naturally.

In the earlier work related to index one saddles in \cite{wwju,ujpyw} the integrable
structure was not emphasized and the uses of the integrals was not developed.
Rather, the emphasis was on the use of the normal form in discovering the phase space structures.
It was hoped that a by-product of the normal form computations
might be useful and intuitive visualization methods
of the kind that have
proved extremely useful and insightful for 2 DoF systems. However, visualization schemes for the
invariant manifolds governing reaction in systems with three and more DoF have thus
far been of limited value.
In \cite{WaalkensBurbanksWigginsb04} attempts were made to visualize the phase space invariant manifolds
governing (planar) HCN isomerization in three dimensions by projecting  the manifolds into the three
dimensional configuration space. In this case, the dividing surface (in a fixed energy surface)
was four dimensional with its equator, the NHIM, three dimensional.  It was difficult to interpret
their projections into the three dimensional configuration space (visually, they did not appear
significantly different).  Moreover, reversibility of Hamilton's equations in this case implies that
the projections of the four dimensional stable and unstable manifolds
into configuration space are identical.  While the visualization of high dimensional phase space structures
is an appealing concept, we believe that the dimensional reduction afforded by integrability of the
normal form in a neighborhood of the equilibrium point is a much more powerful tool for
understanding reactive phenomena. It provides a precise notion of dimensional reduction
to the number of DoFs undergoing ``reactive behaviour'' (i.e. the DoF for which there
is the possibility of
leaving a neighborhood of the equilibrium point). The significance of integrability was
emphasized in \cite{WaalkensSchubertWiggins08} where it was also shown how the phase space
structures can be expressed in terms of the integrals (which is what we have essentially done here).
While their utility for higher dimensional visualization remains to be demonstrated,
for index one saddles the dividing surface plays an important role for a certain types of
sampling \cite{WaalkensWiggins04,WaalkensBurbanksWiggins05,WaalkensBurbanksWiggins05c}
and the NHIM plays a role in determining the flux through the dividing surface.
We will discuss this in more detail in Section \ref{sec:fin_top}.

\subsection{Higher index Saddles}
\label{sec:higher_rank}

Once the generalization from index one to index two saddles has been made,
generalization to higher index saddles is straightforward.
For an $n$ DoF deterministic, time-independent Hamiltonian system,
an index $k$ saddle ($1 \le k \le n$) of the potential energy surface corresponds to
an equilibrium point of Hamilton's  equations where the matrix associated with
the linearization of Hamilton's equations about the equilibrium point has $k$ pairs of real
eigenvalues and $n-k$ pure imaginary pairs of eigenvalues. Assuming that appropriate
generic non-resonance assumptions on the real and imaginary eigenvalues hold, then a
transformation to an integrable Poincar\'e-Birkhoff normal form in  a neighborhood of
the equilibrium point as described above can be carried out. For energies above the equilibrium point,
the system has $k$ integrals describing ``reactive'' dynamics
(i.e., trajectories that can enter and leave a neighborhood of the equilibrium point)
and $n-k$ integrals describing ``bath modes''
(i.e., bounded motions in a neighborhood of the equilibrium point).
Hence integrability allows the reactive dynamics to be studied with a $2k$ dimensional
reduced system with the $k$ reactive integrals providing precise information about
the nature of trajectories that pass through a neighborhood of the equilibrium point.
Moreover, for energies above the saddle, the system has a NHIM of dimension $2n-2k-1$
having $2n-k-1$ dimensional stable and unstable unstable manifolds in the $2n-1$ dimensional energy surface.
In the neighborhood of the equilibrium point,
there are $4^k$ qualitatively distinct trajectory classes.

\subsection{Quantitative Application of Poincar\'e-Birkhoff Normal Form Theory}
\label{sec:nform}

Applying the formalism for higher index saddles
to realistic molecular systems requires software that computes the normal form Hamiltonian,
extracts the integrals of the motion, and computes the coordinate transformation (and its inverse)
from the original physical coordinates to the normal form coordinates, all to sufficiently high order
to yield the desired accuracy (and accuracy must be assessed by a battery of tests
at all points of the calculation). Software to carry out this program for index one saddles is available at
http://lacms.maths.bris.ac.uk/publications/software/index.html.
In principle it is possible to modify this software to carry out the analysis described above for index two,
and higher, saddles.  

\newpage
\section{Index two saddles and isomerization dynamics}
\label{sec:isomerization}

A number of possible contexts in which index two saddles might be important
have already been discussed in Sec.\ \ref{sec:intro}.
It is however useful to specify the
possible dynamical role of index two saddles more precisely.
We consider two examples: first, the dynamical definition of
stepwise (sequential) versus concerted isomerization mechanisms on
a multi-well potential surface; second, the classification of reactive
trajectories in a system in which there are two distinct transition states.
In the latter case our analysis of dynamics in
the vicinity of the index two saddle suggests  a rigorous
identification of so-called ``non-MEP'' \cite{Mann02,Sun02,Debbert02,Ammal03,Lopez07,Lourderaj08}
or ``roaming'' trajectories \cite{Townsend04,Bowman06,Shepler07,Shepler08,Suits08,Heazlewood08}
based on phase space structures.

It should be noted that the partitioning of phase space we discuss is local, that is,
restricted to the vicinity of the index two saddle.

\subsection{Stepwise vs concerted isomerization}

Index two saddles are of dynamical significance
in the identification of sequential versus concerted isomerization mechanisms
in the case of multiple well isomerizations.

\subsubsection{Stepwise versus concerted isomerization in 2 DoF}

\begin{figure}[H]
 \centering
 \includegraphics[width=4.5in]{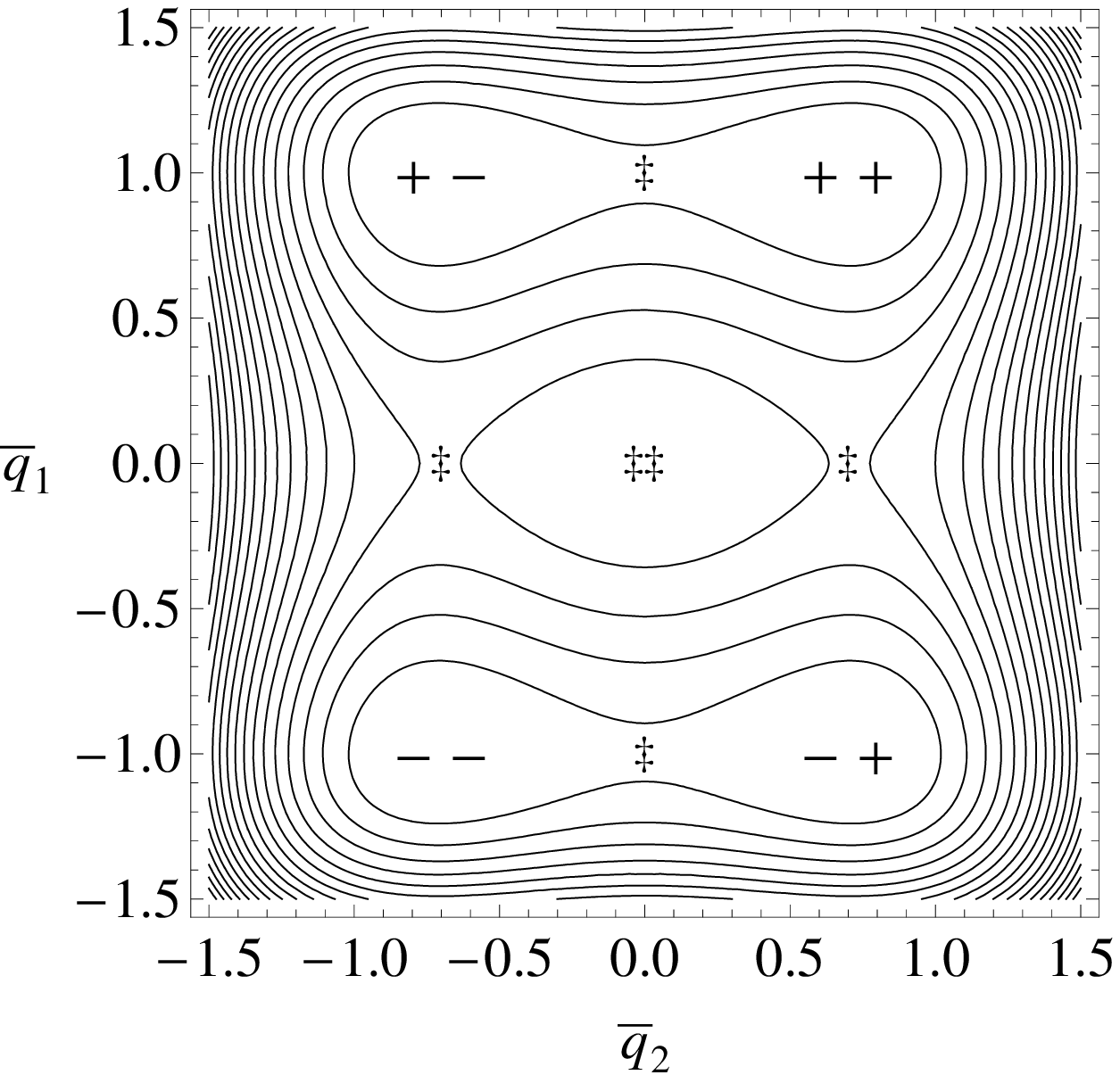}\\
 \caption{\label{plot_4_well} Contour plot of a model 2 DoF potential having 4 minima, 4 index one saddles
 ($\ddagger$) and one index two saddle ($\ddagger\ddagger$).}
 \end{figure}

Consider a system with two coordinates $(\qb_1, \qb_2)$,
having a multiwell potential surface $v(\qb_1, \qb_2)$
of the kind shown schematically
in Figure \ref{plot_4_well} \cite{saddle_footnote6}.
Such a surface can describe, for example,
conformational energies as a function of internal angles \cite{Minyaev95} or
internal rotation/inversion potentials \cite{Minyaev97}.
(See, for example, Figure 1 in ref.\ \onlinecite{Minyaev95}.)

The model potential shown has 4 minima.  Each potential minimum is given a symbolic
label according to the quadrant of the $(\qb_1, \qb_2)$ plane in which it is located.
For example, minimum $(--)$ is located in the quadrant $(\qb_1<0, \qb_2 <0)$.

The minima are connected by index one saddles, each denoted by the symbol $\ddagger$.
At higher energy, in the middle of the potential surface, sits an
index two saddle, denoted $\ddagger \ddagger$ (the ``hilltop'').
(We locate the index two saddle at the coordinate origin, $(\qb_1, \qb_2) = (0, 0)$, and
take the value of the potential at the hilltop to be
zero, $v(0,0) = 0$.)

For total energies $E<0$, the system can only make transitions
between minima by passing through the phase space dividing surfaces
(transition states) associated with the index one saddles (NHIMs).
For such energies, we can define the distinct bound (reactant)
regions of phase space in the usual way, and describe the isomerization
kinetics in terms of the associated phase space volumes and flux
through phase space dividing surfaces associated with the index one saddles.

Classes of isomerizing trajectories associated with particular
transitions between wells can be denoted symbolically: for example,
trajectories of type $(-+;--)$ start in the well $(--)$ and
pass to the well $(-+)$ through the appropriate dividing surface.

It is clear that, for $E<0$, the only way the system can pass from well
$(--)$ to well $(++)$, say, is by a \emph{stepwise} mechanism:
a sequence of isomerizations such as $(+-;--)$ followed by $(++;+-)$
is required.

By contrast, for $E>0$, trajectories are in principle able to ``roam''
around on the potential surface.  In addition to the sequential mechanism just
described, there is another possibility,
namely, a hilltop crossing or \emph{concerted} mechanism, where, for the example just discussed,
a trajectory simply passes ``directly'' from well $(--)$ to well $(++)$
without undergoing the stepwise isomerization process which is necessary
for $E<0$.

It is then natural to ask: how can we
distinguish between these mechanisms in a dynamically rigorous way?
It is not sufficient to analyze trajectories in configuration space, as
both sequentially isomerizing and hilltop crossing trajectories will pass close to the
index two saddle.

The key to dynamical analysis of isomerizing
trajectories is the symbolic trajectory classification introduced
previously. We imagine that the saddle-saddle
normal form has been obtained in the vicinity of the index two saddle.
This means, in particular, that a (nonlinear, in general)
invertible symplectic transformation $T$ has been obtained relating physical
coordinates $(\qb_1, \pb_1, \qb_2, \pb_2)$ and normal form
coordinates $(q_1, p_1, q_2, p_2)$ \cite{ujpyw,WaalkensSchubertWiggins08}
\begin{equation}
(\qb, \pb) = T(q, p).
\end{equation}
In a phase space neighborhood of the saddle-saddle equilibrium, it is therefore
possible to use the inverse transformation $(q, p) = T^{-1}(\qb, \pb)$
to determine the normal form coordinates along a given trajectory.
Given these normal form coordinates and the approximate integrability of
the motion in the vicinity of the index two saddle,
a trajectory with energy $E>0$  can be classified according to the symbolic scheme
developed previously.

For the particular kind of potential surface considered here, the
symbolic classification scheme of trajectories using the normal form
is precisely the desired classification scheme for isomerization
behavior. For example, a trajectory of type $(+-;--)$ passes from well $(--)$ to
well $(+-)$, and so is part of a sequential isomerization process,
while a trajectory of type $(++;--)$ is a hilltop crossing
trajectory that, as determined by a rigorous phase space criterion
in the vicinity of the saddle-saddle equilibrium,
passes directly from well $(--)$ to well $(++)$.

\subsubsection{Stepwise versus concerted isomerization in 3 DoF}

We now introduce a third mode: for simplicity, we first consider
an uncoupled harmonic oscillator DoF.  As discussed in
Sec.\ \ref{sec:sad_sad_cent}, there is now a saddle-saddle-center
equilibrium.
For given total energy $E$, there will be distinct regimes, depending on
the partitioning of energy between mode 3 and the 2 DoF isomerizing subsystem.

If the total energy $E$  is above that of the index two saddle,
but there is sufficient energy in mode 3,
then hilltop crossing isomerization is not possible in the
2 DoF subsystem.  If on
the other hand the amount of energy in mode 3 is small, then the 2 DoF subsystem
can undergo both sequential and hilltop crossing isomerization.
Isomerizing trajectories in the 3 DoF system can then be
classified using the normal form including the center DoF as developed in
Sec.\ \ref{sec:sad_sad_cent}.

Suppose that mode 3 and the 2 DoF subsystem are now coupled.
If we consider the behavior of trajectories initialized
in a region of phase space with a large amount of energy in mode 3
a relevant question is:  what is the mechanism by which the system
will pass from one potential minimum to another, given
that the minima in question are separated by more than one saddle at low energies?

Intuitively, there are two possibilities:
if significant energy transfer between mode 3 and the 2 DoF subsystem
does not occur, then the system simply passes from one well to the other
in a sequential fashion.  If however
energy transfer occurs between mode 3 and
the 2 DoF isomerizing subsystem, it is possible that
sufficient energy can pass into the 2 DoF  subsystem
so that trajectories can isomerize via
the hilltop crossing mechanism.
In this case, hilltop isomerization is mediated by the additional mode.

These dynamical alternatives can in principle be distinguished using
the transformation to normal form coordinates in the
vicinity of the saddle-saddle-center equilibrium (cf.\ Sec.\ \ref{sec:sad_sad_cent}).

\subsection{Competing transition states and non-MEP mechanisms}

We now discuss a second example where analysis of the phase space
structure in the vicinity of an index two saddle is potentially useful
for obtaining deeper mechanistic understanding of an
isomerization reaction.

\begin{figure}[H]
 \centering
 \includegraphics[width=4.5in]{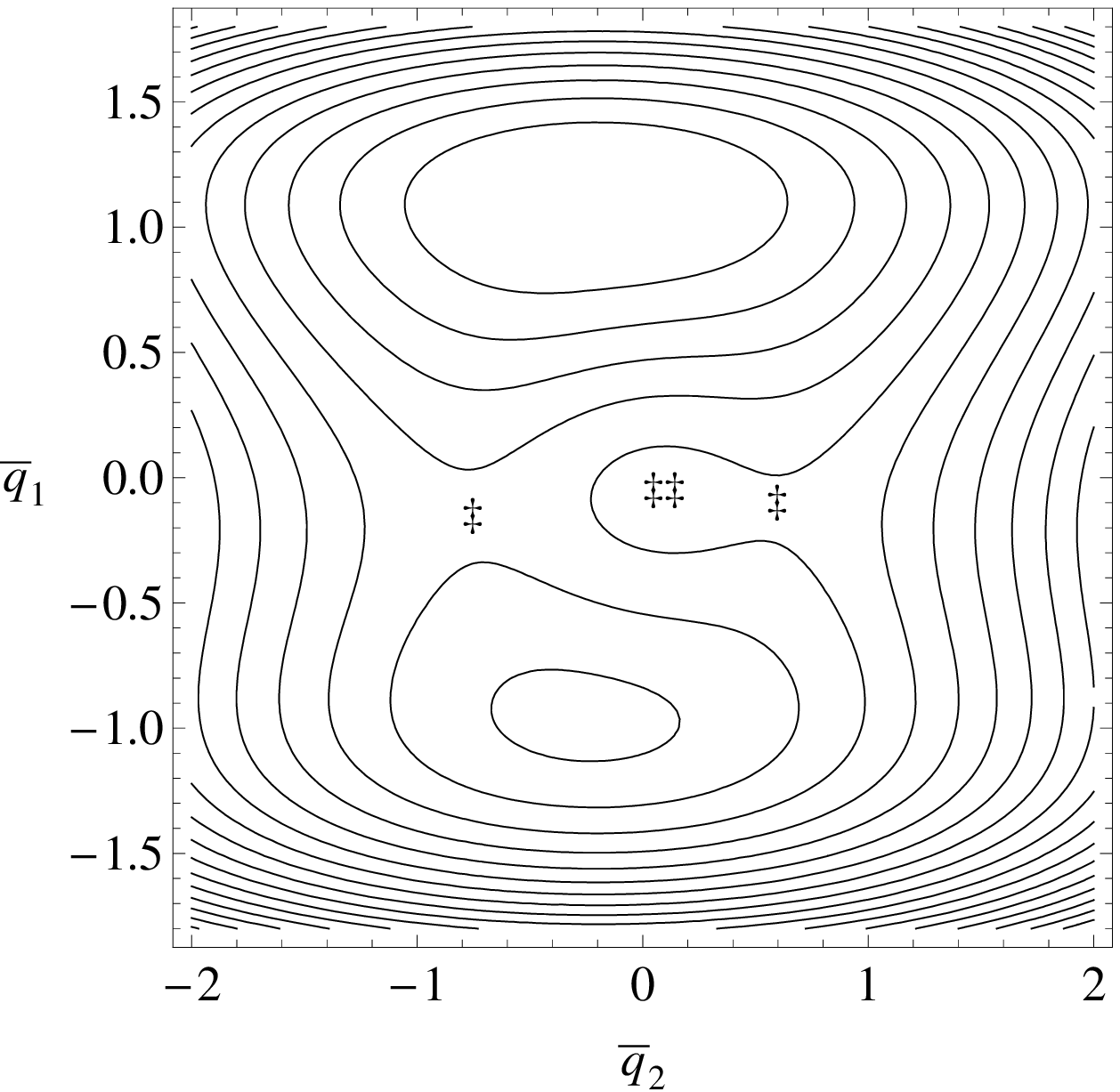}\\
 \caption{\label{plot_2_well} Contour plot of a model 2 DoF potential
 having 2 minima, 2 index one saddles ($\ddagger$)
 and one index two saddle ($\ddagger\ddagger$).}
 \end{figure}

Consider the model 2 DoF potential shown in Figure \ref{plot_2_well}.
This potential has two minima, two index one saddles (having different energies)
and one index two saddle (energy $E=0$).
The index two saddle sits between the two index one saddles
(as per Murrell-Laidler \cite{Murrell68}).
(See, for example, the double proton
transfer reaction in napthazarin, Scheme 14 in ref.\ \onlinecite{Minyaev94}.)

At total energies $E<0$, the system can pass from one well to another
through either of two transition states; in the example shown, the transition state
associated with the index one saddle at $\qb_2 <0$ has lower energy, and hence
is associated with the ``minimum energy path'' mechanism.

At energies above that of the higher energy TS (denoted channel 2), there may exist
some kind of dynamical competition between transition states:
the lower energy TS (channel 1) might be narrow (small flux), while the
higher energy TS (channel 2) might be broader (larger flux).

For energies above the hilltop, $E>0$, there are in principle
three possibilities: trajectories can isomerize via channel 1, via channel 2, or
they can isomerize by ``passsage over the hilltop''.

Once again, the symbolic classification of trajectories in the vicinity
of the saddle-saddle equilibrium enables the
three classes of trajectory to be rigorously identified.
Trajectories of class $(+-;--)$ correspond to the first mechanism (channel 1),
class $(++;-+)$ the second mechanism (channel 2), while trajectories
of class $(+-;-+)$ or $(++;--)$ are associated with
the hilltop crossing isomerization.

All isomerizing trajectories except those of class $(+-;--)$
can therefore be labelled ``non-MEP'' trajectories \cite{Mann02,Sun02,Debbert02,Ammal03,Lopez07,Lourderaj08}.
All such non-MEP trajectories can be regarded as
``roaming'' trajectories \cite{Townsend04,Bowman06,Shepler07,Shepler08,Suits08,Heazlewood08}.

\newpage

\section{Summary and Conclusions}
\label{sec:fin_top}

In this paper we have analyzed the phase space structure in the vicinity of
an equilibrium point associated with an index two saddle or ``hilltop'' on a potential energy
surface (``index two saddle'').  For the case of model quadratic Hamiltonians
we have shown that the behavior of trajectories passing through a
phase space neighborhood of the
equilibrium is fully characterized by the value of associated integrals of the motion.
We have introduced a symbolic classification of trajectories
in the vicinity of the saddle-saddle equilibrium, where the symbolic representation of
trajectory classes is directly related to the nature of the associated hilltop crossing
dynamics.
For the general case where nonlinear terms are present in the Hamiltonian,
Poincar\'e-Birkhoff normal form theory provides a constructive procedure
for obtaining an integrable approximation to the full Hamiltonian in the vicinity of
the equilibrium, allowing a corresponding classification of trajectories to be made
in the general case.

As for the case of index one saddles, normally hyperbolic invariant manifolds (NHIMs)
associated with index two saddles
are an important phase space structure, and we have described
the role they play in phase space transport in the vicinity of
index two saddles.  In particular, we have shown that the normal form transformation
enables co-dimension one surfaces to be defined in phase space;  these surfaces
serve to partition phase space in the vicinity of the
saddle-saddle equilibrium into regions corresponding to distinct trajectory types.

Two examples of the importance of index two saddles in problems of chemical dynamics were considered.
First, isomerization on a potential energy surface with multiple
symmetry equivalent minima;  second, isomerization in a system
having two potential minima connected by two
distinct transition states associated with conventional index one saddles, with
an index two saddle situated between the two index one saddles.
We suggest that classification of different classes of reactive trajectories in the
vicinity of the index two saddle enables a rigorous definition of non-MEP trajectories to
be given in terms of local phase space structure.

To conclude, we list several topics that we believe merit further investigation
for the case of index two saddles.

\begin{description}

\item[Reaction Coordinates.] For the case of index one saddles a natural definition of
a phase space reaction coordinate is obtained by putting all of the energy into
the reactive mode \cite{WaalkensBurbanksWigginsb04,WaalkensSchubertWiggins08}, i.e.,
for energy $H=E$, and the Hamiltonian expressed as a function of the integrals, we set
\begin{equation}
H(I_1, 0, \ldots, 0)=E.
\end{equation}
\noindent
This condition defines a path in the ``reactive plane'' defined by $q_1$ and $p_1$ in
normal form coordinates, which can then be mapped back into the original physical coordinates
by the normal form transformation.

For the case of an index two saddle we have two reactive modes,
described by the integrals $I_1$ and $I_2$, and a four dimensional ``reactive space'' defined
by the coordinates $q_1-p_1-q_2-p_2$.
Can dynamically significant paths analogous to that for the
index one saddle be defined in this reactive space?

\item[Activated Complex.] For index one saddles the NHIM has the interpretation
as the ``activated complex'' \cite{ujpyw,WaalkensSchubertWiggins08}.
Do the NHIMs associated with index $k$ saddles, $1< k \le n$ have a similar chemical significance?

\item[Flux.] For the case of index one saddles the NHIM is the ``boundary'' of the forward
and backward dividing surfaces. The magnitudes of
the fluxes across each of these dividing surfaces are equal and, using
Stokes theorem, are expressed as the integral of an appropriate quantity over the NHIM,
equal to the volume
in the space of ``bath modes'' enclosed by the contour defined by $H(0, I_2, \ldots, I_n) = E$
\cite{WaalkensWiggins04}.

For index two (and higher) saddles it is natural to ask whether or not there
is any chemical significance to the volume in the space of ``bath modes'' enclosed by the contour
defined by $H(0,0,  I_3, \ldots, I_n) = E$.

\item[Gap times and sampling.] Can one develop a useful gap time formalism \cite{Thiele62,Dumont86,Ezra09b}
to characterize rates of passage in and out of phase space regions defined by the co-dimension
one surfaces associated with index two saddles (and higher)?

 Related to the gap time formalism, it has been shown that the
dividing surface provided by the normal form provides a natural surface on which
to sample initial conditions for trajectory calculations that yield quanities of physical
interest such as the reactant volume or dissociation rate \cite{WaalkensBurbanksWiggins05,Ezra09b}.
Does the normal form associated with index two saddles yield manifolds on which a
similar trajectory sampling strategy can be implemented?

\item[The Manifolds $q_1=0$, $p_1=0$ $q_2=0$, and $p_2=0$ , a new type of NHIM?]

We have been careful {\em not} to refer to the co-dimension one invariant manifolds 
defined by $q_1=0$, $p_1=0$ $q_2=0$, and $p_2=0$  as ``NHIMs''.
There are two reasons for this. The first is that they are not of the character of the 
NHIMs associated with index one saddles, i.e., compact and boundaryless
(diffeomorphic to spheres), having (invariant) stable and unstable manifolds (locally) 
partitioning the energy surface into reactants and products,
and  ``persistent under perturbations'' 
(practically, this means  that the manifolds maintain their character for a range of energy above that of the saddle).
The second is that they do {\em not} immediately satisfy the hypotheses of the 
existing  normally hyperbolic invariant manifold theorems \cite{Wiggins94}.
This latter point deserves more discussion.  Roughly speaking, the property of  ``normal hyperbolicity'' 
means that, under linearized dynamics, the magnitude
of the growth and decay of vectors normal to the manifolds dominate 
the magnitude of the growth and  decay of vectors tangent to the manifold.
The manifolds under consideration are obtained by setting one coordinate to zero. 
Therefore, the  growth of vectors under the linearized  dynamics
in the direction of  the coordinate set to zero  describes the  normal dynamics, 
and it is a simple matter to verify that in this direction vectors are
either exponentially growing or decaying (depending on the manifold under consideration). 
The problem is that  there will
also be directions tangent to the manifold that are exponentially growing or decaying.  
Additionally, these manifolds are {\em not} compact and  boundaryless.
Sometimes this  difficulty can be dealt with through a consideration of ``overflowing'' or ``inflowing'' invariance 
(see \cite{Wiggins94} for examples and related references). 
Nevertheless, the  manifolds  do exist for the normal form truncated at {\em any} order, and this is sufficient for our purposes.
Using this property, and the ideas and setting just described, it may be possible to prove something like a ``persistence theorem
for normally hyperbolic invariant manifolds'' for this setting, but we do not pursue that problem here.

\end{description}

Addressing several of these issues will require globalizing the invariant manifolds obtained in the
vicinity of the saddle-saddle equilibribrium.  Indeed, globalization  constitutes an outstanding
challenge for the theory of the dynamics associated with higher index saddles.

\acknowledgments

SW  acknowledges the support of the  Office of Naval Research Grant No.~N00014-01-1-0769.
GSE and SW both
acknowledge the stimulating environment of the NSF sponsored Institute for
Mathematics and its Applications (IMA) at the University of Minnesota,
where this manuscript was completed.

\appendix

\section{Phase Space Geometry and Transport Associated with an $n$ DoF
Quadratic Hamiltonian of a Saddle-Saddle-Center-$\ldots$-Center Equilibrium Point}
\label{sec:ndof_quad}

The following  quadratic Hamiltonian describes the dynamics
near a saddle-saddle-center-$\ldots$-center  stability type equilibrium point
located at the origin:
\begin{equation}
H_{n-quad} = \lambda_1 q_1 p_1 +  \lambda_2 q_2 p_2 +\sum_{i=3}^{n} \frac{\omega_i}{2} \left( q_i^2 + p_i^2 \right),
\label{model_lin_ham}
\end{equation}
\noindent
with associated Hamilton's equations:
\begin{subequations}
\label{hameq_ss_p}
\begin{eqnarray}
\dot{q}_1 & = & \frac{\partial H_{n-quad}}{\partial p_1} =  \lambda_1 q_1,  \\
\dot{p}_1 & = & -\frac{\partial H_{n-quad}}{\partial q_1} =   -\lambda_1 p_1,  \\
\dot{q}_2 & = &  \frac{\partial H_{n-quad}}{\partial p_2} =   \lambda_2 q_2,  \\
\dot{p}_2 & = & -\frac{\partial H_{n-quad}}{\partial q_2}  = -\lambda_2 p_2, \\
\dot{q}_i & = & \frac{\partial H_{n-quad}}{\partial p_i} =   \omega_i p_i,  \quad i=3, \ldots, n\\
\dot{p}_i & = & -\frac{\partial H_{n-quad}}{\partial q_i}  =   -\omega_i q_i,  \quad i=3, \ldots, n.
\end{eqnarray}
\end{subequations}
\noindent
It is straightforward to verify that
$I_1 =q_1p_1, \, I_2 = q_2 p_2, \, I_k = \frac{\omega_k}{2} \left( q_k^2 + p_k^2 \right), \, k=3, \ldots, n$
are  (independent) integrals of the motion for \eqref{hameq_ss_p}.
As previously, we define a neighborhood of the saddle-saddle-center equilibrium point in phase space as follows:
\begin{equation}
{\cal L} \equiv \left\{ (q_1, p_1, \ldots, q_n, p_n) \, \mid \, \vert q_i \vert \le  \varepsilon_i, \, \vert p_i \vert \le  \varepsilon_i, \, i=1, \ldots, n
\right\},
\label{calL}
\end{equation}
\noindent
for appropriately chosen $\varepsilon_i >0$, $i=1, \ldots, n$.
Again, the fact that Hamilton's equations are separable,
and that the motion of $q_k$ and $p_k$ is bounded, $k=3, \ldots, n$
means that the issue of trajectories entering $\cal L$ and exiting $\cal L$ depends
entirely on the behavior of the $(q_1, p_1, q_2, p_2)$ components of a trajectory and,
as a consequence of separability,
this behavior is exactly as described in Section \ref{sec:sad_sad}.

As for the 3 DoF case considered in Sec.\ \ref{sec:3dof},  the equilibrium point is located
at the origin and has zero energy,
and we will  consider geometric structures in the energy surface for positive energies:
\begin{equation}
H_{n-quad} = \lambda_1 q_1 p_1 +  \lambda_2 q_2 p_2 +
\sum_{k=3}^{n}\frac{\omega_k}{2} \left( q_k^2 + p_k^2 \right)=E>0.
\label{E_surf_ndof}
\end{equation}
\noindent
It is clear from  \eqref{hameq_ss_p} that $q_1=p_1=q_2=p_2=0$ is a
two dimensional invariant manifold in the $2n$ dimensional phase space.
Using \eqref{E_surf_ndof}, its intersection with the $2n-1$ dimensional energy surface is given by:

\begin{equation}
\sum_{k=3}^{n}\frac{\omega_k}{2} \left( q_k^2 + p_k^2 \right)=E>0.
\label{nhim_ndof}
\end{equation}

\noindent
This is the equation for a $2n-5$ dimensional sphere, $S^{2n-5}$, in the energy surface;
as above, it is an example of a NHIM \cite{Wiggins90,Wiggins94}.
The coordinates $q_1, p_1, q_2, p_2$ can be viewed as coordinates for
the normal directions of the NHIM, and it follows
from the nature of the asymptotic (in time) behavior of these coordinates (see \eqref{hameq_ss_p})
that this NHIM has a $2n-3$ dimensional stable manifold and a $2n-3$ dimensional
unstable manifold in the $2n-1$ dimensional energy surface.

\def\cprime{$'$}

\newpage

\newpage

\newpage

\newpage

\begin{thebibliography}{10}

\bibitem{Wigner38}
E.~P. Wigner.
\newblock {The transition state method}.
\newblock {\em Trans. Faraday Soc.}, 34:29--40, 1938.

\bibitem{Keck67}
J.~C. Keck.
\newblock {Variational theory of reaction rates}.
\newblock {\em Adv. Chem. Phys.}, XIII:85--121, 1967.

\bibitem{Pechukas81}
P.~Pechukas.
\newblock {Transition state theory}.
\newblock {\em Ann. Rev. Phys. Chem.}, 32:159--177, 1981.

\bibitem{Truhlar83}
D.~G. Truhlar, W.~L. Hase, and J.~T. Hynes.
\newblock {Current Status of Transition-State Theory}.
\newblock {\em J. Phys. Chem.}, 87:2664--2682, 1983.

\bibitem{Anderson95}
J.~B. Anderson.
\newblock {Predicting Rare Events in Molecular Dynamics}.
\newblock {\em Adv. Chem. Phys.}, XCI:381--431, 1995.

\bibitem{Truhlar96}
D.~G. Truhlar, B.~C. Garrett, and S.~J. Klippenstein.
\newblock {Current Status of Transition-State Theory}.
\newblock {\em J. Phys. Chem.}, 100:12711--12800, 1996.

\bibitem{saddle_footnote1}
{Consider a potential energy function $V=V(q_1, \ldots, q_n)$ that is a
  function of $n$ coordinates $\{q_k\}$. (Coordinates describing translation
  and rotation are excluded.) At a non-degenerate critical point of $V$, where
  $\partial V/\partial q_k =0$, $k=1,\ldots,n$, the Hessian matrix $\partial^2
  V/\partial q_i\partial q_j$ has $n$ nonzero eigenvalues. The \emph{index} of
  the critical point is the number of negative eigenvalues.}

\bibitem{Wiggins90}
S.~Wiggins.
\newblock On the geometry of transport in phase space {I}. {T}ransport in
  $k$-degree-of-freedom {H}amiltonian systems, $2 \le k < \infty$.
\newblock {\em Physica D}, 44:471--501, 1990.

\bibitem{wwju}
S.~Wiggins, L.~Wiesenfeld, C.~Jaffe, and T.~Uzer.
\newblock Impenetrable barriers in phase-space.
\newblock {\em Phys. Rev. Lett.}, 86(24):5478--5481, 2001.

\bibitem{ujpyw}
T.~Uzer, C.~Jaffe, J.~Palacian, P.~Yanguas, and S.~Wiggins.
\newblock The geometry of reaction dynamics.
\newblock {\em Nonlinearity}, 15:957--992, 2002.

\bibitem{WaalkensBurbanksWiggins04}
H.~Waalkens, A.~Burbanks, and S.~Wiggins.
\newblock A computational procedure to detect a new type of high-dimensional
  chaotic saddle and its application to the 3{D} {H}ill's problem.
\newblock {\em J. Phys. A}, 37:L257--L265, 2004.

\bibitem{WaalkensWiggins04}
H.~Waalkens and S.~Wiggins.
\newblock Direct construction of a dividing surface of minimal flux for
  multi-degree-of-freedom systems that cannot be recrossed.
\newblock {\em J. Phys. A}, 37:L435--L445, 2004.

\bibitem{WaalkensBurbanksWigginsb04}
H.~Waalkens, A.~Burbanks, and S.~Wiggins.
\newblock Phase space conduits for reaction in multidimensional systems: {HCN}
  isomerization in three dimensions.
\newblock {\em J. Chem. Phys.}, 121(13):6207--6225, 2004.

\bibitem{WaalkensBurbanksWiggins05}
H.~Waalkens, A.~Burbanks, and S.~Wiggins.
\newblock Efficient procedure to compute the microcanonical volume of initial
  conditions that lead to escape trajectories from a multidimensional potential
  well.
\newblock {\em Physical Review Letters}, 95:084301, 2005.

\bibitem{WaalkensBurbanksWiggins05c}
H.~Waalkens, A.~Burbanks, and S.~Wiggins.
\newblock A formula to compute the microcanonical volume of reactive initial
  conditions in transition state theory.
\newblock {\em J. Phys. A}, 38:L759--L768, 2005.

\bibitem{SchubertWaalkensWiggins06}
R.~Schubert, H.~Waalkens, and S.~Wiggins.
\newblock Efficient computation of transition state resonances and reaction
  rates from a quantum normal form.
\newblock {\em Phys. Rev. Lett.}, 96:218302, 2006.

\bibitem{WaalkensSchubertWiggins08}
H.~Waalkens, R.~Schubert, and S.~Wiggins.
\newblock {W}igner's dynamical transition state theory in phase space:
  Classical and quantum.
\newblock {\em Nonlinearity}, 21(1):R1--R118, 2008.

\bibitem{MacKay90}
R.~S. MacKay.
\newblock {Flux over a saddle}.
\newblock {\em Phys. Lett. A}, 145:425--427, 1990.

\bibitem{Komatsuzaki00}
T.~Komatsuzaki and R.~S. Berry.
\newblock {Local regularity and non-recrossing path in transition state: a new
  strategy in chemical reaction theories}.
\newblock {\em J. Mol. Struct. THEOCHEM}, 506:55--70, 2000.

\bibitem{Komatsuzaki02}
T.~Komatsuzaki and R.~S. Berry.
\newblock {Chemical reaction dynamics: Many-body chaos and regularity}.
\newblock {\em Adv. Chem. Phys.}, 123:79--152, 2002.

\bibitem{Wiesenfeld03}
L.~Wiesenfeld, A.~Faure, and T.~Johann.
\newblock {Rotational transition states: relative equilibrium points in
  inelastic molecular collisions}.
\newblock {\em J. Phys. B}, 36:1319--1335, 2003.

\bibitem{Wiesenfeld04}
L.~Wiesenfeld.
\newblock {Local separatrices for Hamiltonians with symmetries}.
\newblock {\em J. Phys. A}, 37:L143--L149, 2004.

\bibitem{Wiesenfeld04a}
L.~Wiesenfeld.
\newblock {Dynamics with a rotational transition state}.
\newblock {\em Few Body Syst.}, 34:163--168, 2004.

\bibitem{Komatsuzaki05}
T.~Komatsuzaki, K.~Hoshino, and Y.~Matsunaga.
\newblock {Regularity in Chaotic Transitions on Multi-Basin Landscapes}.
\newblock {\em Adv. Chem. Phys.}, 130 B:257--313, 2005.

\bibitem{Jaffe05}
C.~Jaff\'{e}, K.~Shinnosuke, J.~Palacian, P.~Yanguas, and T.~Uzer.
\newblock {A New Look at the Transition State: Wigner's Dynamical Perspective
  Revisited}.
\newblock {\em Adv. Chem. Phys.}, 130 A:171--216, 2005.

\bibitem{Wiesenfeld05}
L.~Wiesenfeld.
\newblock {Geometry of phase space transition states: many dimensions, angular
  momentum}.
\newblock {\em Adv. Chem. Phys.}, 130 A:217--265, 2005.

\bibitem{Gabern05}
F.~Gabern, W.~S. Koon, J.~E. Marsden, and S.~D. Ross.
\newblock {Theory and computation of non-RRKM lifetime distributions and rates
  in chemical systems with three or more degrees of freedom}.
\newblock {\em Physica D}, 211:391--406, 2005.

\bibitem{Gabern06}
F.~Gabern, W.~S. Koon, J.~E. Marsden, and S.~D. Ross.
\newblock {Application of tube dynamics to non-statistical reaction processes}.
\newblock {\em Few-Body Systems}, 38:167--172, 2006.

\bibitem{Shojiguchi08}
A.~Shojiguchi, C.~B. Li, T.~Komatsuzaki, and M.~Toda.
\newblock {Dynamical foundation and limitations of statistical reaction
  theory}.
\newblock {\em Comm. Nonlinear Sci. Numerical Simulation}, 13:857--867, 2008.

\bibitem{saddle_footnote2}
{Without loss of generality, throughout this paper all equilibrium points
  considered will be located at the origin with zero energy.}

\bibitem{Heidrich86}
D.~Heidrich and W.~Quapp.
\newblock {Sadddle points of index 2 on potential energy surfaces and their
  role in theoretical reactivity investigations}.
\newblock {\em Theor. Chim. Acta}, 70:89--98, 1986.

\bibitem{Mezey87}
P.~G. Mezey.
\newblock {\em {Potential Energy Hypersurfaces}}.
\newblock Elsevier, Amsterdam, 1987.

\bibitem{Wales03}
D.~J. Wales.
\newblock {\em {Energy Landscapes}}.
\newblock Cambridge University Press, Cambridge, 2003.

\bibitem{Jensen98}
F.~Jensen.
\newblock {Transition structure optimization techniques}.
\newblock In P.~V.~R. Schleyer, editor, {\em {Encyclopedia of Computational
  Chemistry}}, pages 3114--3123. Wiley, New York, 1998.

\bibitem{Minyaev91}
R.~M. Minyaev.
\newblock {Molecular structure and global description of the potential energy
  surface}.
\newblock {\em J. Struct. Chem.}, 32:559--589, 1991.

\bibitem{Murrell68}
J.~N. Murrell and K.~J. Laidler.
\newblock {Symmetries of activated complexes}.
\newblock {\em Trans. Faraday Soc.}, 64:371--377, 1968.

\bibitem{Wales92}
D.~J. Wales and R.~S. Berry.
\newblock {Limitations of the Murrell-Laidler theorem}.
\newblock {\em J. Chem. Soc. Faraday Trans.}, 88:543--544, 1992.

\bibitem{Minyaev94}
R.~M. Minyaev.
\newblock {Reaction path as a gradient line on a potential energy surface}.
\newblock {\em Int. J. Quantum Chem.}, 49:105--127, 1994.

\bibitem{Minyaev94a}
R.~M. Minyaev.
\newblock {Gradient lines on multidimensional potential energy surfaces and
  chemical reaction mechanisms}.
\newblock {\em Russ. Chem. Rev.}, 63:883--903, 1994.

\bibitem{Minyaev04}
R.~M. Minyaev, I.~V. Getmanskii, and W.~Quapp.
\newblock {A second-order saddle point in the reaction coordinate for the
  isomerization of the NH5 complex: Ab initio calculations}.
\newblock {\em Russ. J. Phys. Chem.}, 78:1494--1498, 2004.

\bibitem{Minyaev95}
R.~M. Minyaev.
\newblock {Correlated internal rotations of CH$_3$ groups in dimethyl ether and
  BH$_2$ groups in diborylmethylene}.
\newblock {\em Russ. J. Phys. Chem.}, 69:1463--1471, 1995.

\bibitem{Minyaev97}
R.~M. Minyaev and E.~A. Lepin.
\newblock {Internal rotation in the H$_2$BNH$_2$ molecule}.
\newblock {\em Russ. J. Phys. Chem.}, 71:1449--1453, 1997.

\bibitem{Fau95}
S.~Fau and G.~Frenking.
\newblock {Anti van't Hoff/Le Bel geometries of boron compounds. A theoretical
  study of classical and non-classical isomers of B$_2$CH$_4$, B$_2$NH$_3$ and
  B$_2$OH$_2$}.
\newblock {\em Theochem J. Mol. Struct.}, 338:117--130, 1995.

\bibitem{Salem71}
L.~Salem.
\newblock {Narcissistic reactions: synchronism vs nonsynchronism in
  automerizations and enatiomerizations}.
\newblock {\em Acc. Chem. Res.}, 4:322--328, 1971.

\bibitem{Carpenter04}
B.~K. Carpenter.
\newblock {Potential energy surfaces and reaction dynamics}.
\newblock In R.~A. Moss, M.~S. Platz, and M.~Jones Jr., editors, {\em {Reactive
  Intermediate Chemistry}}, pages 925--960. Wiley, New York, 2004.

\bibitem{Meroueh02}
S.~O. Meroueh, Y.~F. Wang, and W.~L. Hase.
\newblock {Direct dynamics simulations of collision and surface-induced
  dissociation of N-protonated glycine. Shattering fragmentation}.
\newblock {\em J. Phys. Chem. A}, 106:9983--9992, 2002.

\bibitem{Cavagna01}
A.~Cavagna, I.~Giardini, and G.~Parisi.
\newblock {Role of saddles in mean-field dynamics above the glass transition}.
\newblock {\em J. Phys. A}, 34:5317--5326, 2001.

\bibitem{Cavagna01a}
A.~Cavagna.
\newblock {Fragile vs. strong liquids: A saddles-ruled scenario}.
\newblock {\em Europhys. Lett.}, 53:490--496, 2001.

\bibitem{Doye02}
J.~P.~K. Doye and D.~J. Wales.
\newblock {Saddle points and dynamics of Lennard-Jones clusters, solids, and
  supercooled liquids}.
\newblock {\em J. Chem. Phys.}, 116:3777--3788, 2002.

\bibitem{Wales03a}
D.~J. Wales and J.~P.~K. Doye.
\newblock {Stationary points and dynamics in high-dimensional systems}.
\newblock {\em J. Chem. Phys.}, 119:12409--12416, 2003.

\bibitem{Angelini03}
L.~Angelini, G.~Ruocco, and F.~Zamponi.
\newblock {Saddles and dynamics in a solvable mean-field model}.
\newblock {\em J. Chem. Phys.}, 118:8301--8306, 2003.

\bibitem{Shell04}
M.~S. Shell, P.~G. Debenedetti, and A.~Z. Panagiotopoulos.
\newblock {Saddles in the energy landscape: Extensivity and thermodynamic
  formalism}.
\newblock {\em Phys. Rev. Lett.}, 92:035506, 2004.

\bibitem{Grigera06}
T.~S. Grigera.
\newblock {Geometrical properties of the potential energy of the soft-sphere
  binary mixture}.
\newblock {\em J. Chem. Phys.}, 124:064502, 2006.

\bibitem{Coslovich07}
D.~Coslovich and G.~Pastore.
\newblock {Understanding fragility in supercooled Lennard-Jones mixtures. II.
  Potential energy surface}.
\newblock {\em J. Chem. Phys.}, 127:124505, 2007.

\bibitem{Angelini08}
L.~Angelini, G.~Ruocco, and F.~Zamponi.
\newblock {Role of saddles in topologically driven phase transitions: The case
  of the d-dimensional spherical model}.
\newblock {\em Phys. Rev. E}, 77:052101, 2008.

\bibitem{Kastner08}
M.~Kastner.
\newblock {Phase transitions and configuration space topology}.
\newblock {\em Rev. Mod. Phys.}, 80:167--187, 2008.

\bibitem{Mann02}
D.~J. Mann and W.~L. Hase.
\newblock {Ab initio direct dynamics study of cyclopropyl radical
  ring-opening}.
\newblock {\em J. Am. Chem. Soc.}, 124:3208--3209, 2002.

\bibitem{Sun02}
L.~P. Sun, K.~Y. Song, and W.~L. Hase.
\newblock {A S(N)2 reaction that avoids its deep potential energy minimum}.
\newblock {\em Science}, 296:875--878, 2002.

\bibitem{Debbert02}
S.~L. Debbert, B.~K. Carpenter, D.~A. Hrovat, and W.~T. Borden.
\newblock {The iconoclastic dynamics of the 1,2,6-heptatriene rearrangement}.
\newblock {\em J. Am. Chem. Soc.}, 124:7896--7897, 2002.

\bibitem{Ammal03}
S.~C. Ammal, H.~Yamataka, M.~Aida, and M.~Dupuis.
\newblock {Dynamics-driven reaction pathway in an intramolecular
  rearrangement}.
\newblock {\em Science}, 299:1555--1557, 2003.

\bibitem{Lopez07}
J.~G. Lopez, G.~Vayner, U.~Lourderaj, S.~V. Addepalli, S.~Kato, W.~A. Dejong,
  T.~L. Windus, and W.~L. Hase.
\newblock {A direct dynamics trajectory study of F-+CH(3)OOH reactive
  collisions reveals a major Non-IRC reaction path}.
\newblock {\em J. Am. Chem. Soc.}, 129:9976--9985, 2007.

\bibitem{Lourderaj08}
K.~Park U.~Lourderaj and W.~L. Hase.
\newblock {Classical trajectory simulations of post-transition state dynamics}.
\newblock {\em Int. Rev. Phys. Chem.}, 27:361--403, 2008.

\bibitem{Townsend04}
D.~Townsend, S.~A. Lahankar, S.~K. Lee, S.~D. Chambreau, A.~G. Suits, X.~Zhang,
  J.~Rheinecker, L.~B. Harding, and J.~M. Bowman.
\newblock {The roaming atom: Straying from the reaction path in formaldehyde
  decomposition}.
\newblock {\em Science}, 306:1158--1161, 2004.

\bibitem{Bowman06}
J.~M. Bowman.
\newblock {Skirting the transition state, a new paradigm in reaction rate
  theory}.
\newblock {\em PNAS}, 103:16061--16062, 2006.

\bibitem{Shepler07}
B.~C. Shepler, B.~J. Braams, and J.~M. Bowman.
\newblock {Quasiclassical trajectory calculations of acetaldehyde dissociation
  on a global potential energy surface indicate significant non-transition
  state dynamics}.
\newblock {\em J. Phys. Chem. A}, 111:8282--8285, 2007.

\bibitem{Shepler08}
B.~C. Shepler, B.~J. Braams, and J.~M. Bowman.
\newblock {``Roaming'' dynamics in CH3CHO photodissociation revealed on a
  global potential energy surface}.
\newblock {\em J. Phys. Chem. A}, 112:9344--9351, 2008.

\bibitem{Suits08}
A.~G. Suits.
\newblock {Roaming atoms and radicals: A new mechanism in molecular
  dissociation}.
\newblock {\em Acc. Chem. Res.}, 41:873--881, 2008.

\bibitem{Heazlewood08}
B.~R. Heazlewood, M.~J.~T. Jordan, S.~H. Kable, T.~M. Selby, D.~L. Osborn,
  B.~C. Shepler, B.~J. Braams, and J.~M. Bowman.
\newblock {Roaming is the dominant mechanism for molecular products in
  acetaldehyde photodissociation}.
\newblock {\em PNAS}, 105:12719--12724, 2008.

\bibitem{Maxwell54}
J.~C. Maxwell.
\newblock {\em {A Treatise on Electricity and Magnetism}}, volume~1.
\newblock Dover, New York, 1954.

\bibitem{haller_cite}
{G.\ Haller, J.\ Palacian, P.\ Yanguas, T.\ Uzer and C.\ Jaff\'{e}, \emph{Comm.
  Nonlinear Sci. Num. Simul.}, to appear, 2009}.

\bibitem{Wiggins94}
S.~Wiggins.
\newblock {\em Normally hyperbolic invariant manifolds in dynamical systems}.
\newblock Springer-Verlag, 1994.

\bibitem{saddle_footnote3}
{Briefly, the co-dimension of a manifold is the dimension of the space in which
  the manifold exists, minus the dimension of the manifold. The significance of
  a manifold being ``co-dimension one'' is that it is one less dimension than
  the space in which it exists. Therefore it can ``divide'' the space and act
  as a separatrix, or barrier, to transport. Co-dimension two manifolds do not
  have this property.}

\bibitem{saddle_footnote4}
{For the quadratic Hamiltonians we consider $\varepsilon_1$ (and
  $\varepsilon_2$) can be chosen ${\cal O}(1)$. When Hamiltonians with higher
  order terms are considered a more careful consideration of the size of
  $\varepsilon_1$ and $\varepsilon_2$ is required.}

\bibitem{qm_footnote}
{The notational convention used here is inspired by quantum mechanics: if the
  initial (incident) state is denoted $i$, the final (scattered) state denoted
  $f$, the quantum mechanical scattering amplitude is given by the matrix
  element $\langle f \vert \hat{S}\vert i \rangle$, with $\hat{S}$ the
  scattering operator. The transition associated with the scattering event is
  then $f \leftarrow i$.}

\bibitem{normal_form_footnote1}
{The normal form for the saddle-saddle-center-$\ldots$-center has not received
  a great deal of individual attention in the literature. Nevertheless, this
  particular form of the normal form has been well-known for some time. See,
  for example, refs \onlinecite{eliasson,dn}. See also the documentation of the
  normal form software available from
  \url{http://lacms.maths.bris.ac.uk/publications/software/index.html}.}

\bibitem{saddle_footnote5}
{The normal form algorithm operates in an iterative fashion by simplifying
  terms in the Taylor expansion about the equilibrium point ``order by order'',
  i.e. the order $M$ terms are normalizes, then the order $M+1$ terms are
  normalized, etc. The algorithm is such that normalization at order $M$ does
  not affect any of the lower order terms (which have already been normalized).
  The point here is that although the algorithm can be carried out to
  arbitrarily high order, practically we must stop the normalization at some
  order (i.e. truncate the Hamiltonian). In practice we have found that
  truncations of the normal form Hamiltonian can be extremely accurate
  according to various tests which we discuss in Section \ref{sec:accuracy}.}

\bibitem{WaalkensBurbanksWiggins05b}
H.~Waalkens, A.~Burbanks, and S.~Wiggins.
\newblock Escape from planetary neighborhoods.
\newblock {\em Mon. Not. R. Astron. Soc.}, 361:763--775, 2005.

\bibitem{Pollak78}
E.~Pollak and P.~Pechukas.
\newblock {Transition states, trapped trajectories, and bound states embedded
  in the continuum}.
\newblock {\em J. Chem. Phys.}, 69:1218--1226, 1978.

\bibitem{saddle_footnote6}
{We assume a kinetic energy of standard form, separable and quadratic in the
  conjugate momenta $(\pb_1, \pb_2)$.}

\bibitem{Thiele62}
E.~Thiele.
\newblock Comparison of the classical theories of unimolecular reactions.
\newblock {\em J. Chem. Phys.}, 36(6):1466--1472, 1962.

\bibitem{Dumont86}
R.~S. Dumont and P.~Brumer.
\newblock {Dynamic theory of statistical unimolecular decay}.
\newblock {\em J. Phys. Chem.}, 90:3509--3516, 1986.

\bibitem{Ezra09b}
{G. S. Ezra, H. Waalkens and S. Wiggins, Microcanonical rates, gap times, and
  phase space dividing surfaces, arXiv:0901.2721v1 [cond-mat.stat-mech]}.

\bibitem{eliasson}
H.~Eliasson.
\newblock {Normal forms for Hamiltonian systems with Poisson commuting
  integrals--elliptic case}.
\newblock {\em Comment. Math. Helvetici}, 65:4--35, 1990.

\bibitem{dn}
H.~Dullin and S.~{V}u Ngoc.
\newblock Symplectic invariants near hyperbolic-hyperbolic points.
\newblock {\em Regular and Chaotic Dynamics}, 12(6):689--716, 2007.

\end{thebibliography}
\end{document}